\documentclass[12pt]{ar}
\usepackage[english]{babel}
\usepackage{psfig,graphicx}

\onecolumn

\renewcommand{\degr} {{$\rm ^o$}}

\begin{document}

\title{High--latitude supergiants: anomalies in the spectrum of LN\,Hya in 2010}

\author{Valentina Klochkova \& Vladimir Panchuk}

\institute{Special Astrophysical Observatory RAS, Nizhnij Arkhyz,  Russia} 

\date{\today}	     

\abstract{High-resolution echelle spectra taken with the 6-m telescope in
2003--2011 are used to study features of the optical spectrum and the
velocity field in the atmosphere of the semiregular variable LN~Hya in
detail. The weak symmetric photospheric absorptions indicate radial
velocity variations from night to night (by as much as 3\,km/s), resulting
from small pulsations. Peculiarities and profile variations were found for
strong lines of Fe\,I, Fe\,II, Ba\,II, Si\,II, etc. The profiles of these
lines were asymmetric: their short-wave wings were extended and
their cores were either split or distorted by emission. During the 2010
observing season, the position and depth of the H$\alpha$ absorption
component, the intensities of the shortand long-wave emission components,
and the intensity ratio of the latter components varied from spectrum to
spectrum. Weak emissions of neutral atoms (VI, MnI, CoI, NiI, FeI)
appeared in the spectrum of June 1, 2010. All these spectral peculiarities,
recorded for the first time, suggest that we have detected rapid changes
in the physical conditions in the upper atmospheric layers of LN\,Hya in
2010.}

\authorrunning{\it Klochkova \& Panchuk}
\titlerunning{\it Spectrum of LN Hya in 2010}

\maketitle

\section{Introduction}

The semiregular variable LN\,Hya (HD\,112374, IRAS\,12538$-$2611) is
located at a high Galactic latitude (b\,=\,36$\lefteqn{.}$\degr4),
indicating that the star belongs to an old Galactic population. Its
brightness variation amplitude is $\Delta$B\,=\,0.32$^{\rm m}$ ; the
variability period is not firmly established, and possible durations range
from 44 to 86 days [1]. Semiregular variables are far evolved post-AGB
stars. Stars of intermediate masses (with initial masses of $3\div
8{\mathcal M}_{\odot}$) are observed in this short-lived stage; they lose
matter via their stellar winds as they evolve of the AGB. The secular
variability of the principal parameters of post--AGB stars has stimulated
spectroscopic monitoring. Such spectroscopic monitoring of selected AGB
and post-AGB stars has been performed with the 6-m telescope of the Special
Astrophysical Observatory over the past decade. The main aim of our
monitoring is to detect peculiarities and possible variations of the
spectrum, and to study the time behavior of the velocity field in the
stellar atmospheres and shells. Observations on the 6-m telescope have
detected variations in the spectra of V510\,Pup [2, 3], V2324\,Cyg [4],
and the optical counterpart of the IR--ray source IRAS\,01005\,+\,7910
[5]. New results concerning variations of the optical spectra and
atmospheric velocity fields have also been obtained for the variable stars
QY\,Sge [6], V354\,Lac [7, 8], V448\,Lac [9], CY\,CMi [10].

Stellar evolution from the AGB stage to the planetary-nebula stage remains
incompletely understood. This concerns the mechanisms and details of the
stellar mass loss and the complex morphology of their circumstellar
gaseous-dusty shells. The role of pulsations and binarity has also been
poorly studied, and the detection of these effects for stars with complex
atmospheric velocity fields is laborious and time consuming. These
considerations also apply to studies of the variable supergiant LN\,Hya.
An image taken with the Hubble Space Telescope [11] shows LN\,Hya as a
point-like object. The star is surrounded with a gas and dust shell, whose
presence in the optical is revealed by the emission profile of the
H$\alpha$ line. After publication of the IRAS results and optical
identification of IRAS objects via catalog cross correlation, LN\,Hya was
identified with the infrared source IRAS\,12538$-$2611. However, the
fluxes in the IRAS bands are low [12] compared to the star's radiation in
the visible. Note that low IR--fluxes have also been observed for several
peculiar UU\,Her-type supergiants at high Galactic latitudes. LN\,Hya is
usually considered to be a member of this rare Galactic population (e.g.
[13, 14]). No infrared excess was detected for the high-latitude
supergiant UU\,Her, whose properties will also be discussed below.

The brightest star in its class, LN\,Hya has been studied spectroscopically
many times. Already in the early 1980s, Luck et al. [15] determined the
main parameters (effective temperature Teff\,=\,6000\,K, surface gravity
log\,g\,=\,0.4 and 0.8) and detailed chemical composition of LN\,Hya, which
displays a metal deficiency, [Fe/H]\,=$-1.2$, and large nitrogen excess,
[N/Fe]\,=\,+0.5. Somewhat later, Klochkova and Panchuk [16] used
photographic spectra taken with the main stellar spectrograph of the 6-m
telescope to study the spectrum of LN\,Hya among a sample of supergiants
at high Galactic latitudes, and confirmed the main conclusion of Luck et
al. [15]: LN\,Hya is a post-AGB star with a deficiency of heavy elements.
Ten years later, Giridhar~et~al.~[17] obtained new estimates of the
fundamental parameters and chemical composition of LN\,Hya based on CCD
spectra, reaching similar conclusions. The main focus of the recent study
[18] was abundances of elements with low condensation temperatures, which
are not subject to selective separation processes (C, N, O, S, Zn).

Thus, several papers devoted to the chemical composition of
LN\,Hya have been published over the last 30 years, but very little
information is available about the peculiarities of its spectrum or
possible time variability. We have obtained new high-quality optical
spectra of LN\,Hya over the last decade. This paper presents our results
on variability and previously unknown peculiarities of the star's spectrum
and its detailed radial-velocity pattern.

\begin{table}[t]
\caption{Log of observations and the heliocentric radial velocities
               Vr of LN\,Hya}
\medskip                                              
\begin{tabular}{ c| c | c | c | r c}                 
\hline
Date    &    JD 2450000+ & $\Delta\lambda$\,,\AA{} &\multicolumn{3}{c}{Vr, km/s}           \\
\cline{4-6}
        &      &  &Vr(metals)& Vr(H$\alpha$$_{\rm abs}$)& Vr(NaI$_{\rm emis}$) \\
\hline
21.02.2003 &  2692.52 &     5150--6660 &$-$23.20$\pm$0.12 (125)&$-$25.5 &$-$21.9  \\
02.04.2010 &  5289.39 &     5160--6680 &$-$23.68$\pm$0.08 (304)&$-$20.5 &$-$21.1  \\
05.04.2010 &  5292.33 &     5160--6680 &$-$21.93$\pm$0.07 (312)&$-$20.7 &$-$22.0  \\
01.06.2010 &  5349.27 &     5220--6690 &$-$28.22$\pm$0.11 (211)&$-$7.8  &$-$21.6  \\
12.01.2011 &  5574.59 &     5210--6680 &$-$23.30$\pm$0.06 (235)&$-$22.3 &$-$21.8  \\
14.03.2011 &  5635.42 &     5220--6690 &$-$24.88$\pm$0.07 (153)&$-$23.6 &$-$21.6  \\
\hline
\end{tabular}
\label{date}
\end{table}

Vr(metals) is the mean velocity from symmetric absorption features; the
number of lines used to derived the mean velocity is given in brackets.
Vr(H$\alpha$$_{\rm abs}$) is the velocity from the H$\alpha$ absorption
component, and Vr(Na$_{\rm emis}$) the velocity from the emission components
of the D--lines NaI.

\section{Obseravtions, reduction, and spectral analysis}

We taken spectroscopic data on LN\,Hya at the Nasmyth focus of the 6-meter
telescope using the NES echelle spectrograph [19, 20]. These observations
used a 2048$\times$2048-pixel CCD and an image slicer [20]. The
spectroscopic resolution was R$\ge$60000 and the S/N ratio S/N$>$100. To
increase the time interval covered by our observations of LN\,Hya, we added
a spectrum with resolution R\,=\,25000 we had earlier obtained at the
Nasmyth focus of the 6-m telescope using the ``Lynx'' echelle spectrograph
[21]. The mean times of our observations (JD) and the recorded spectral
ranges are presented in Table\,\ref{date}.

We extracted 1D spectra from the 2D echelle frames using modified version
[22] of the ECHELLE context of the MIDAS package. To remove cosmic-ray
traces, we performed median averaging of two successive spectra. Our
wavelength calibration were done using spectra from a ThAr hollow cathode
lamp. We checked the instrumental agreement between the stellar and lamp
spectra using telluric [OI], O$_2$, and H$_2$O lines. A more detailed
description of the procedure used to measure radial velocities Vr from
spectra obtained with the NES spectrograph and various sources of
uncertainty can be found in [23]. The rms uncertainty of the Vr
measurements for stars with narrow absorption lines is $<$1.0 km/s (from
{\bf a single} line).

\begin{figure}[t]	      		      
\includegraphics[angle=-90,width=0.95\textwidth,bb=40 60 540 790,clip]{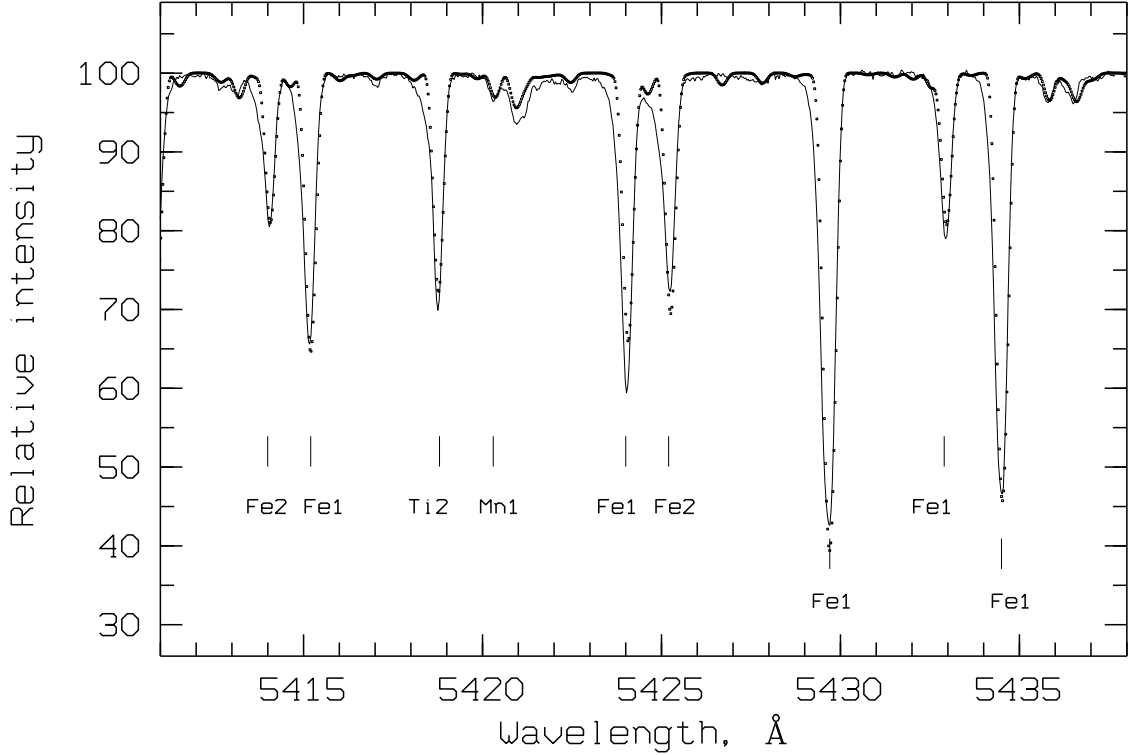}
\caption{Fragment of the spectrum of LN Hya observed on January 12, 2011,
         compared to the theoretical spectrum (points) calculated for
         Teff\,=\,6000\,K, log\,g\,=\,1.0, $\xi_t = 4.7$. Identifications
         of the main spectral lines in this range are indicated.}
\label{synth}
\end{figure}

In addition to position measurements, a multifaceted analysis of the
observed spectra of LN\,Hya requires computations of synthetic spectra. We
applied the SynthV code developed by Tsymbal [24], adapted by Tsymbal to
run on a personal computer in OS~Linux environment. Note that the effective
temperatures Teff derived for LN Hya in different studies vary over a
fairly wide range, 5500$\div$6300 K. We adopted temperature close to the
average of these, Teff\,=\,6000 K, taking into account the good agreement
of the observed spectrum and the theoretical spectrum for this
temperature. Figure\,\ref{synth} displays a part of the spectrum observed
on January 12, 2011 that contains absorption lines almost undistorted by
emission, compared to the theoretical spectrum. Although the computations
and observations are in general agreement, this figure demonstrates that
the short-wavelength absorption wings are extended, even during the
quiescent atmospheric state of LN\,Hya.

\section{Main results}

\subsection{Emission components and line-profile variations}

{\bf The H$\alpha$ profile.} LN\,Hya was classified as an F3\,Ia
supergiant based on its optical spectrum [25]. The first studies dealing
with analyses of the star's spectrum noted the presence of emission in the
H$\alpha$ profile (e.g., [15]). Emission components in the H$\alpha$ are
characteristic of post-AGB stars (for examples, see Fig.\,2 in [26]), and
indeed represent one of the main selection criteria for such objects [27].
Spectra of UU\,Her--type stars also display complex H$\alpha$ emission and
absorption profiles. The H$\alpha$ in the spectrum of the prototype,
UU\,Her, contains two emission components of variable intensity [28]. The
main characteristics of UU\,Her and LN\,Hya are similar in whole. Both
stars are semiregular F--supergiants at high Galactic latitudes. The
elemental abundances in their atmospheres, studied in [28] for UU\,Her and
in [16, 17] for LN\,Hya, are also similar: their metallicities are
[Fe/H]\,=$-$1.32 and $-$1.2, respectively, with large excesses of nitrogen
and deficiency of carbon and heavy s-process metals. Significant
differences between the two stars include, first, the considerably larger
brightness-variation amplitude of UU\,Her compared to LN Hya ($>2^{\rm
m}$), and second, the IR--excess of LN\,Hya, which is absent for UU\,Her.
In addition to the IR--excess, the presence of dust in the shell of LN\,Hya
is revealed by the polarization of the star's light [29].

\begin{figure}[t]	      		      
\hbox{
\includegraphics[angle=0,width=0.35\textwidth,bb=40 60 540 790,clip]{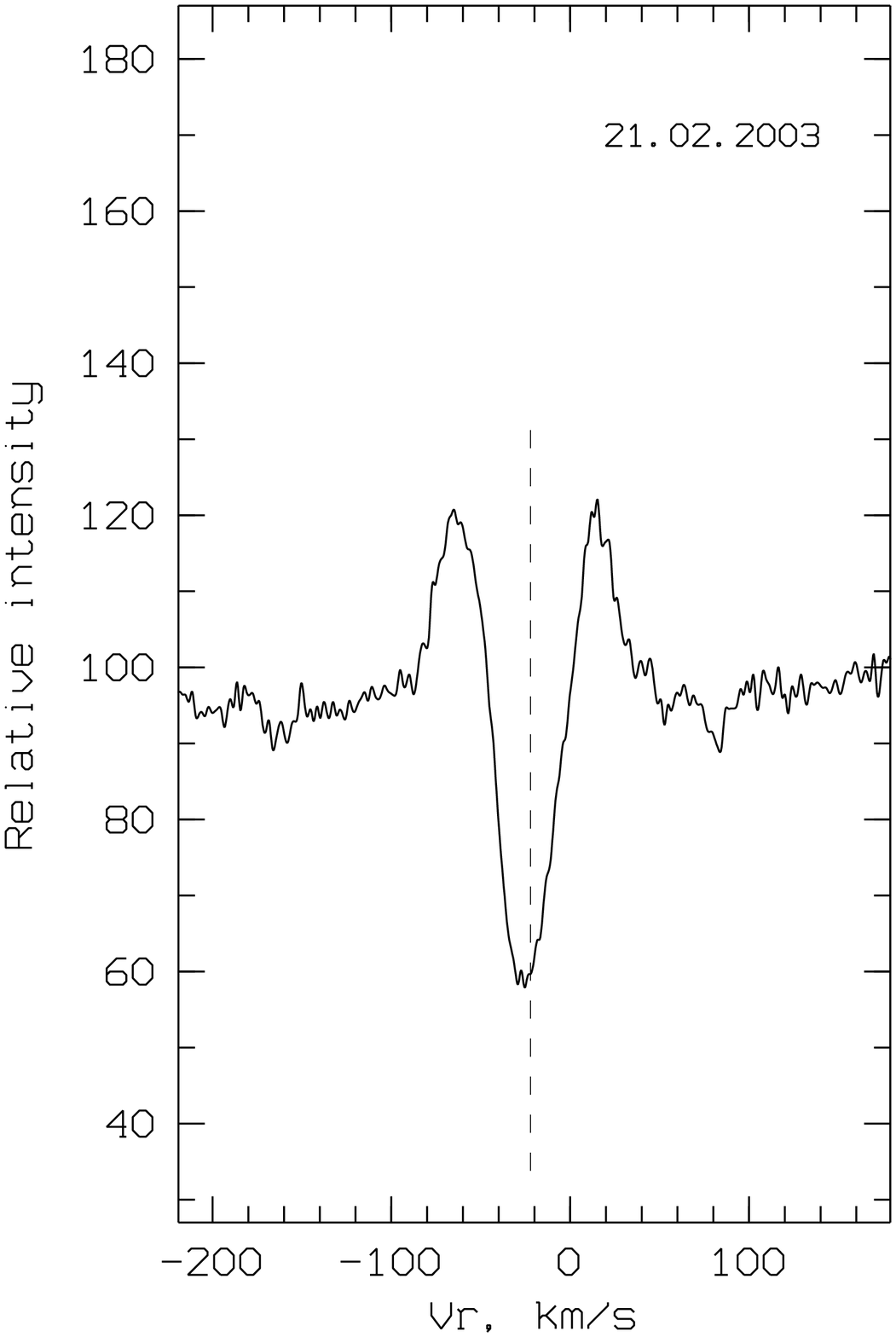}
\includegraphics[angle=0,width=0.35\textwidth,bb=40 60 540 790,clip]{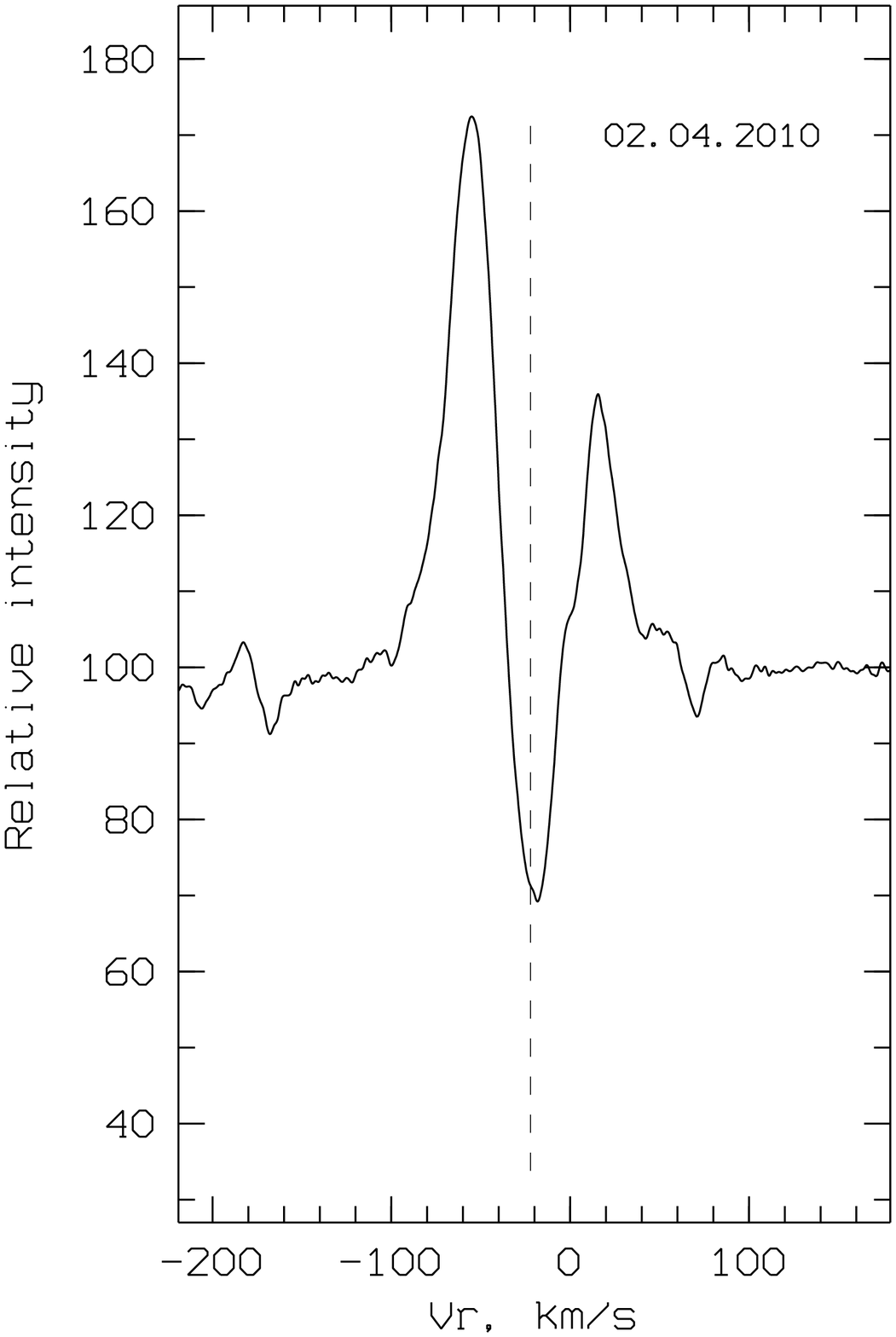}
\includegraphics[angle=0,width=0.35\textwidth,bb=40 60 540 790,clip]{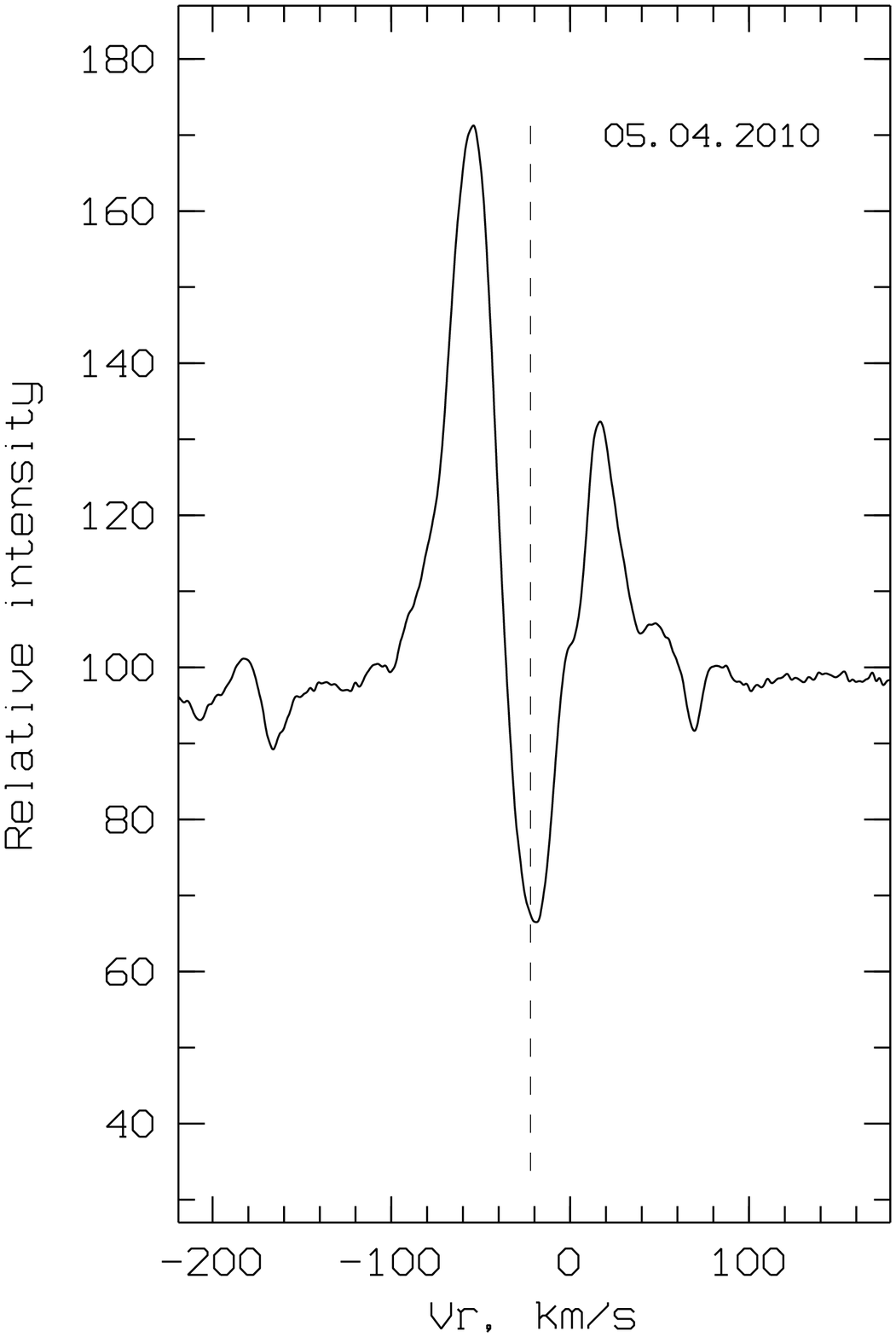}}
\hbox{
\includegraphics[angle=0,width=0.35\textwidth,bb=40 60 540 790,clip]{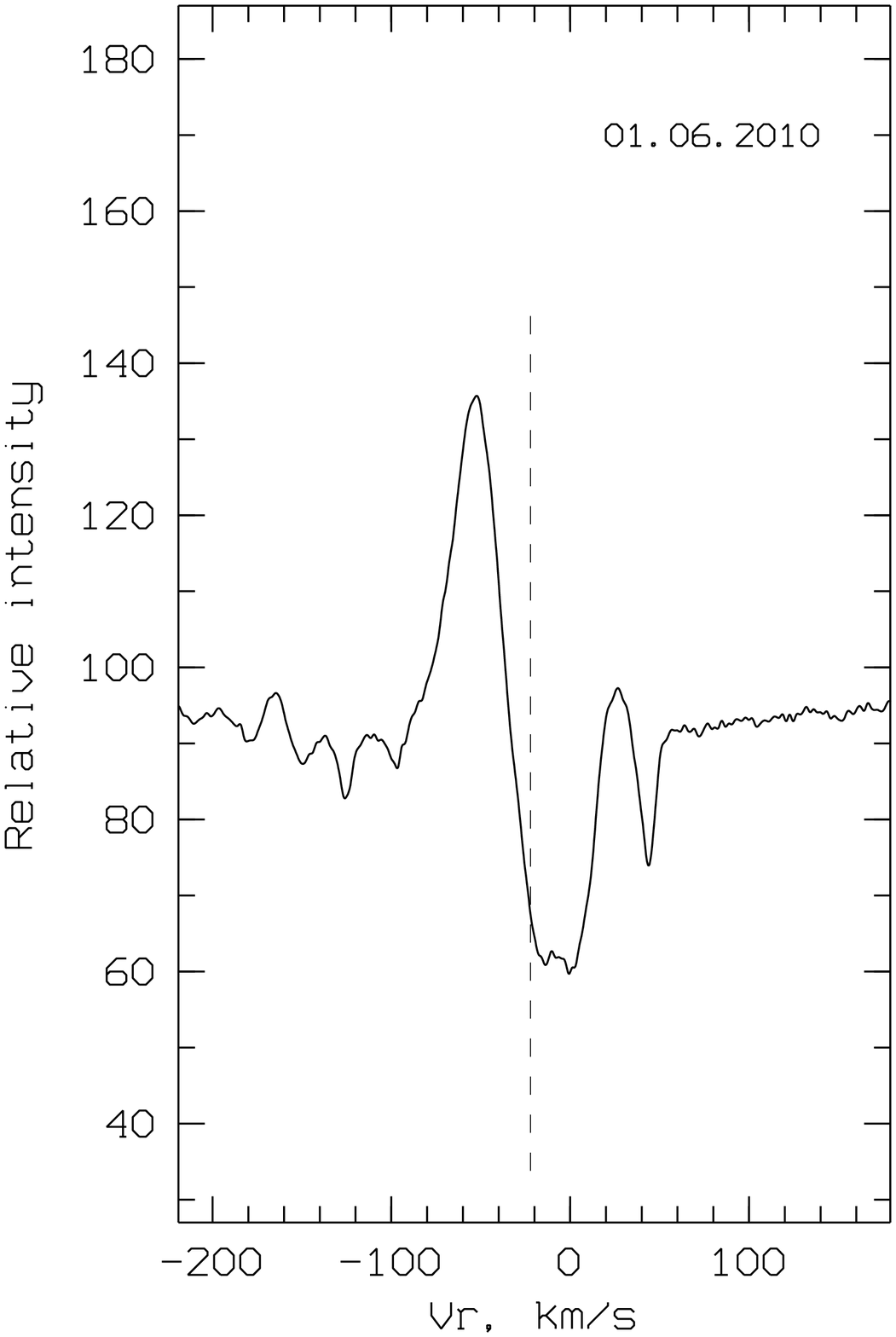}
\includegraphics[angle=0,width=0.35\textwidth,bb=40 60 540 790,clip]{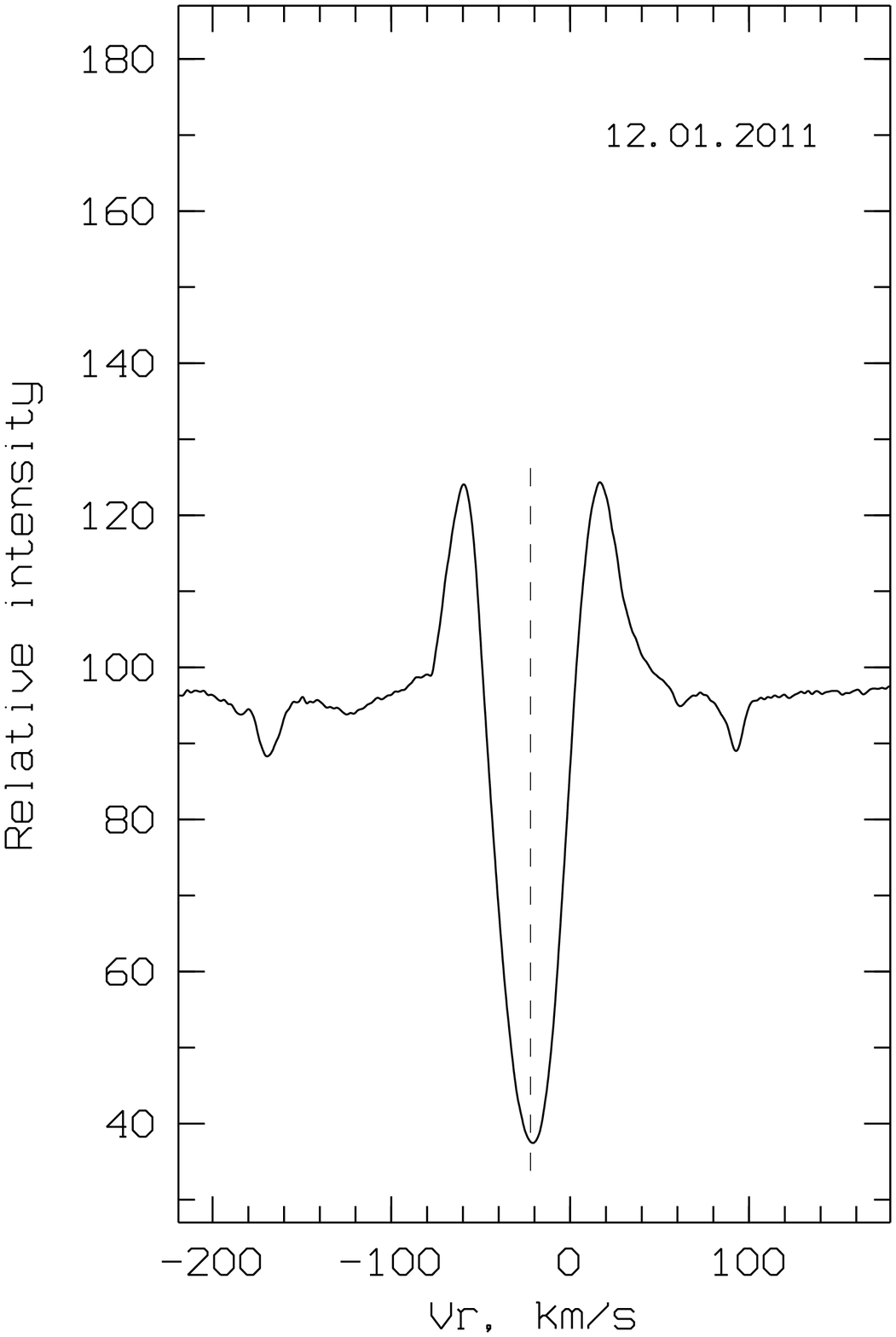}
\includegraphics[angle=0,width=0.35\textwidth,bb=40 60 540 790,clip]{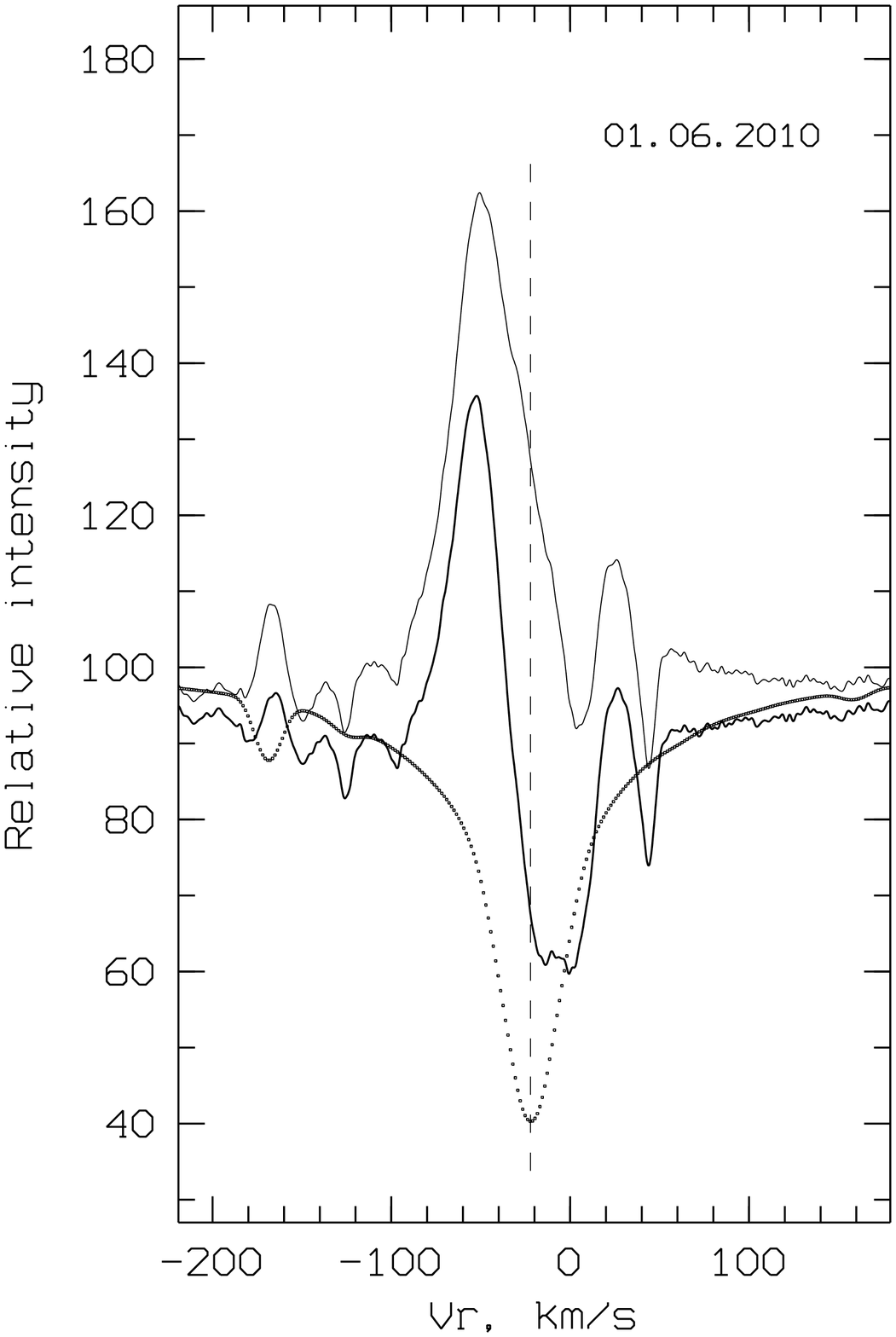}}
\caption{Profies of the H$\alpha$ line in the spectra of LN\,Hya taken on
         various dates from Table\,\ref{date}. In addition to the profile for 1 June, 2010,
         the bottom right panel presents the theoretical H$\alpha$ profile for
         Teff\,=\,6000\,K, log\,g\,=\,1.0, and $\xi_t = 4.7$\,km/s (dotted curve),
         together with the difference between the observed and theoretical
         spectra (thin curve). The vertical dashed lines mark the adopted systemic
         velocity, Vsys\,=$-$21.6\,km/s.}
\label{Halpha}
\end{figure}

The emission in the H$\alpha$ line was long believed to be the only
spectral peculiarity of LN\,Hya. Figure\,\ref{Halpha} shows that all our
spectra of LN\,Hya demonstrate H$\alpha$ with emission components. The
H$\alpha$ profile with two--peaked emission, typical of post--AGB stars,
that was observed in 2003 and 2011 is in agreement with the properties of
the profile in the considerably earlier study of Luck et al. [15]. A large
number of observations of post--AGB stars near the H$\alpha$ line were
obtained, analyzed, and classified by Sanchez Contreras et al. [30]. Using
their classification, we find that the H$\alpha$ profile in the spectrum of
LN\,Hya during its quiescent phases is an emission--filled absorption
(EFA) profile. It is usually believed that this type of the profile
indicates the presence of a long-lived reservoir of circumstellar gas
(rotating disk). The line width is determined primarily by scattering on
free electrons and the kinematics of the circumstellar structure. The
central H$\alpha$ absorption is formed by the peripheral zones of the same
structure, which are at rest relative to the system's center of mass.

We detected strong variations of the H$\alpha$ profile in the spectra of
LN\,Hya obtained in the 2010 observing season. Figure\,\ref{Halpha} shows
changes in the position of the absorption core, the intensities of both
emission components, and their intensity ratio. In addition, during
observations of the least ``excited'' spectra (21 February, 2003, 12
January, 2011, and 14 March, 2011), the H$\alpha$ possessed extended
absorption wings formed in deep layers of the stellar atmosphere, along
with emission components of approximately the same intensity.
Figure\,\ref{Halpha} compares the H$\alpha$ profile observed on 1 June,
2010 and the theoretical profile calculated using the star's fundamental
parameters. The profile in this spectrum differs considerably from those
observed during quiescent phases, and can be considered an inverse P\,Cygni
type profile. In this case, the position of the H$\alpha$ absorption
component differed considerably from its position at other epochs. In
addition, the H$\alpha$ core was considerably shifted towards longer
wavelengths relative to the symmetric metallic absorption lines (by about
15\,km/s).

\begin{figure}[t]	      		      
\hbox{
\includegraphics[angle=0,width=0.35\textwidth,bb=40 60 540 790,clip]{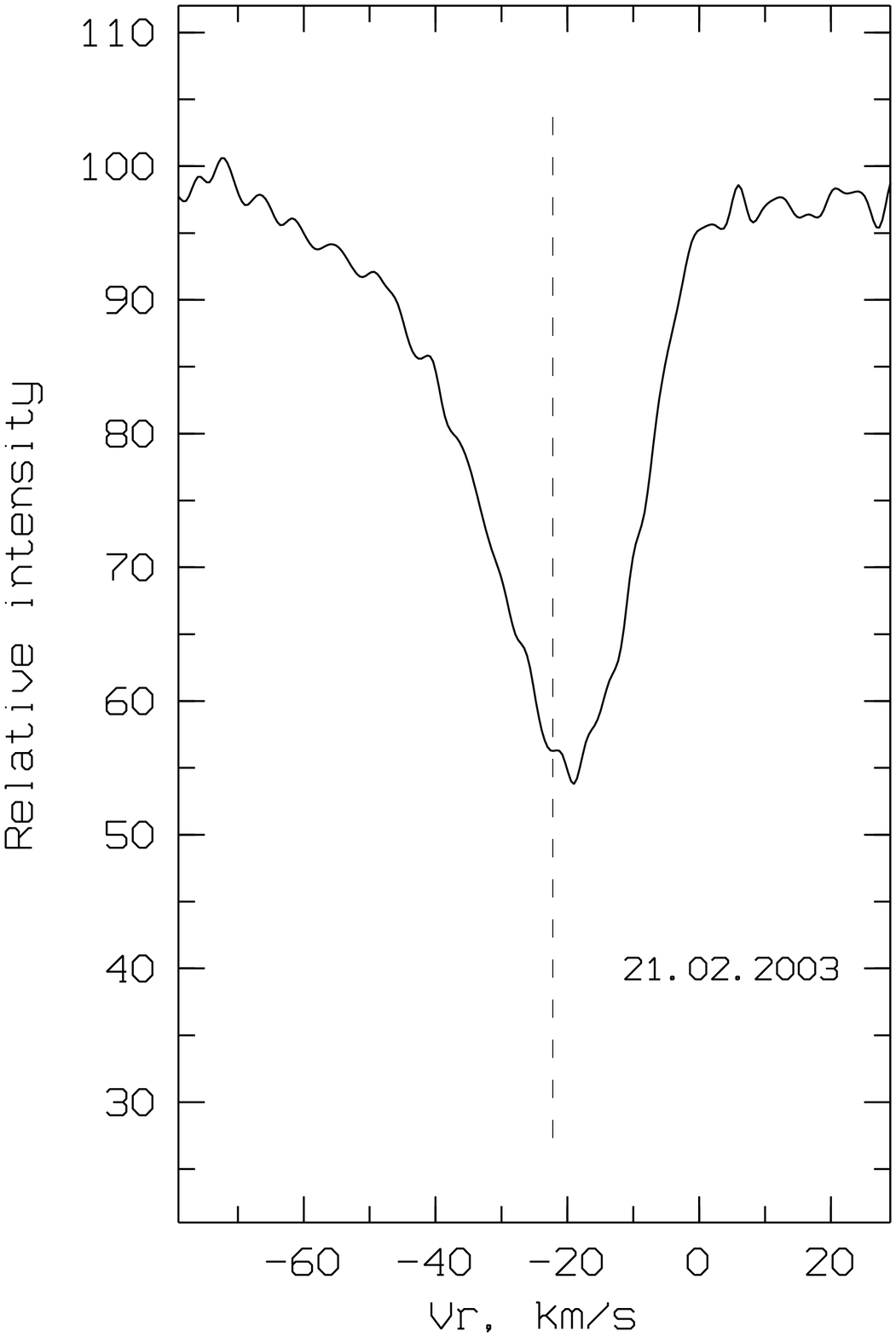}
\includegraphics[angle=0,width=0.35\textwidth,bb=40 60 540 790,clip]{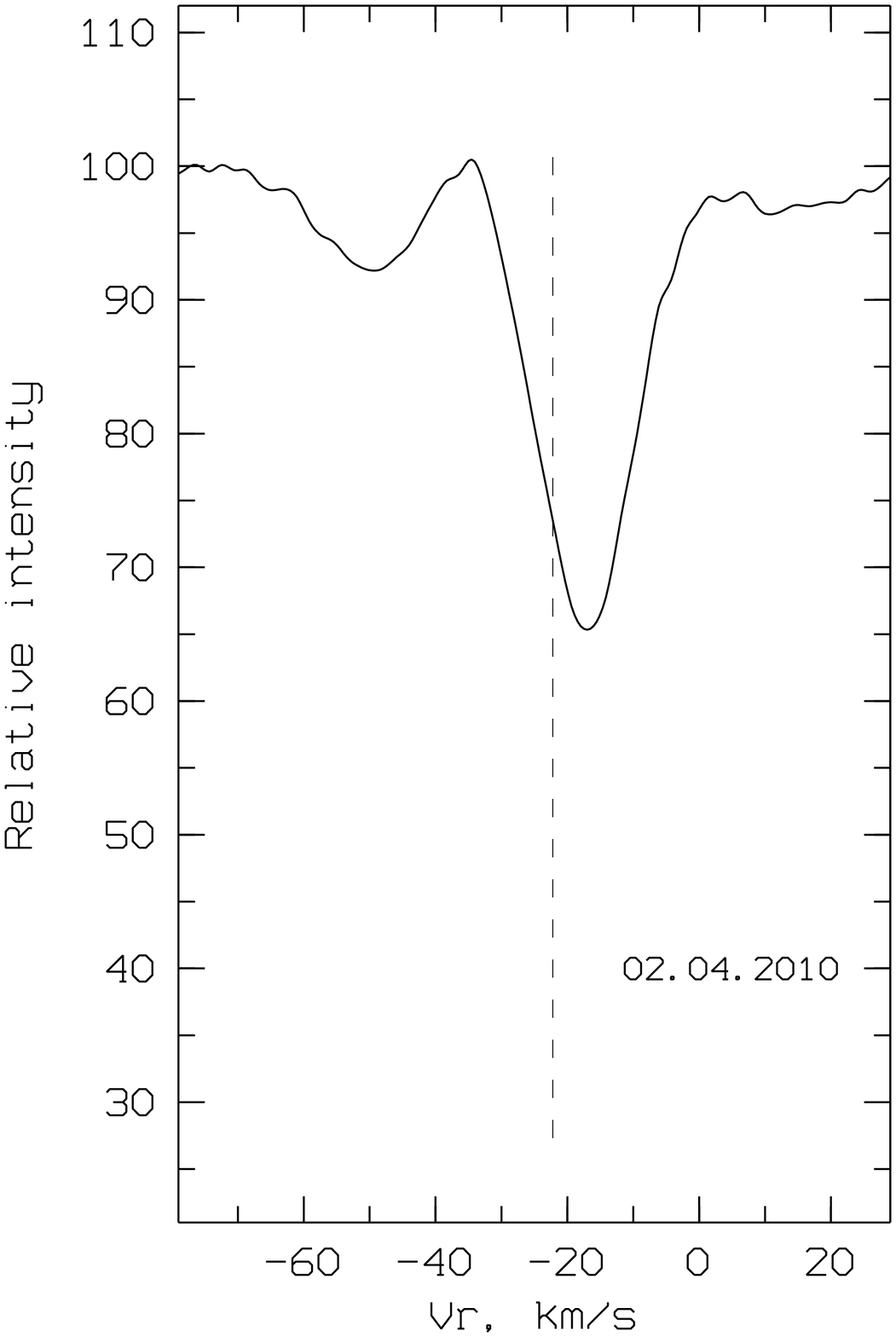}
\includegraphics[angle=0,width=0.35\textwidth,bb=40 60 540 790,clip]{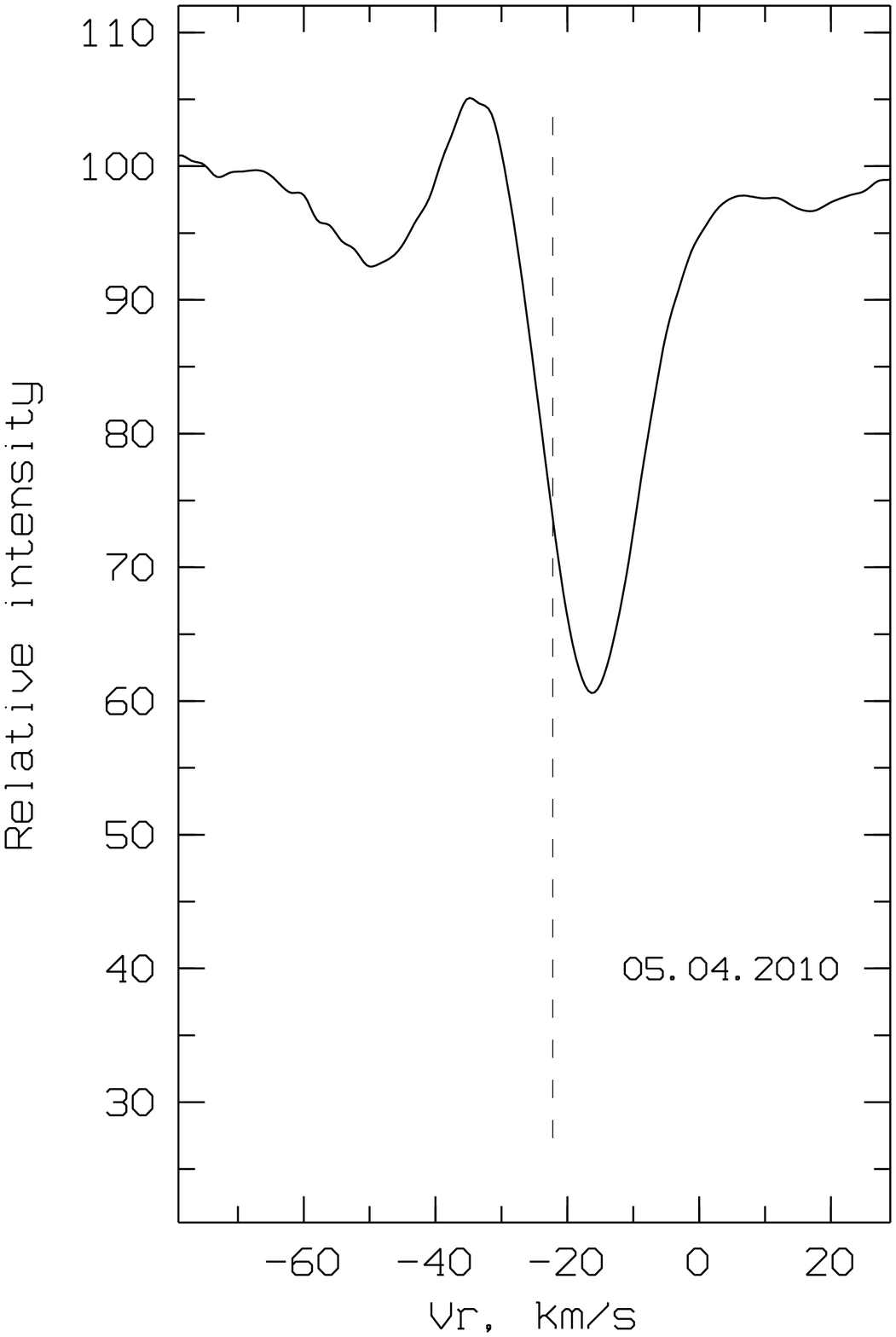}}
\hbox{
\includegraphics[angle=0,width=0.35\textwidth,bb=40 60 540 790,clip]{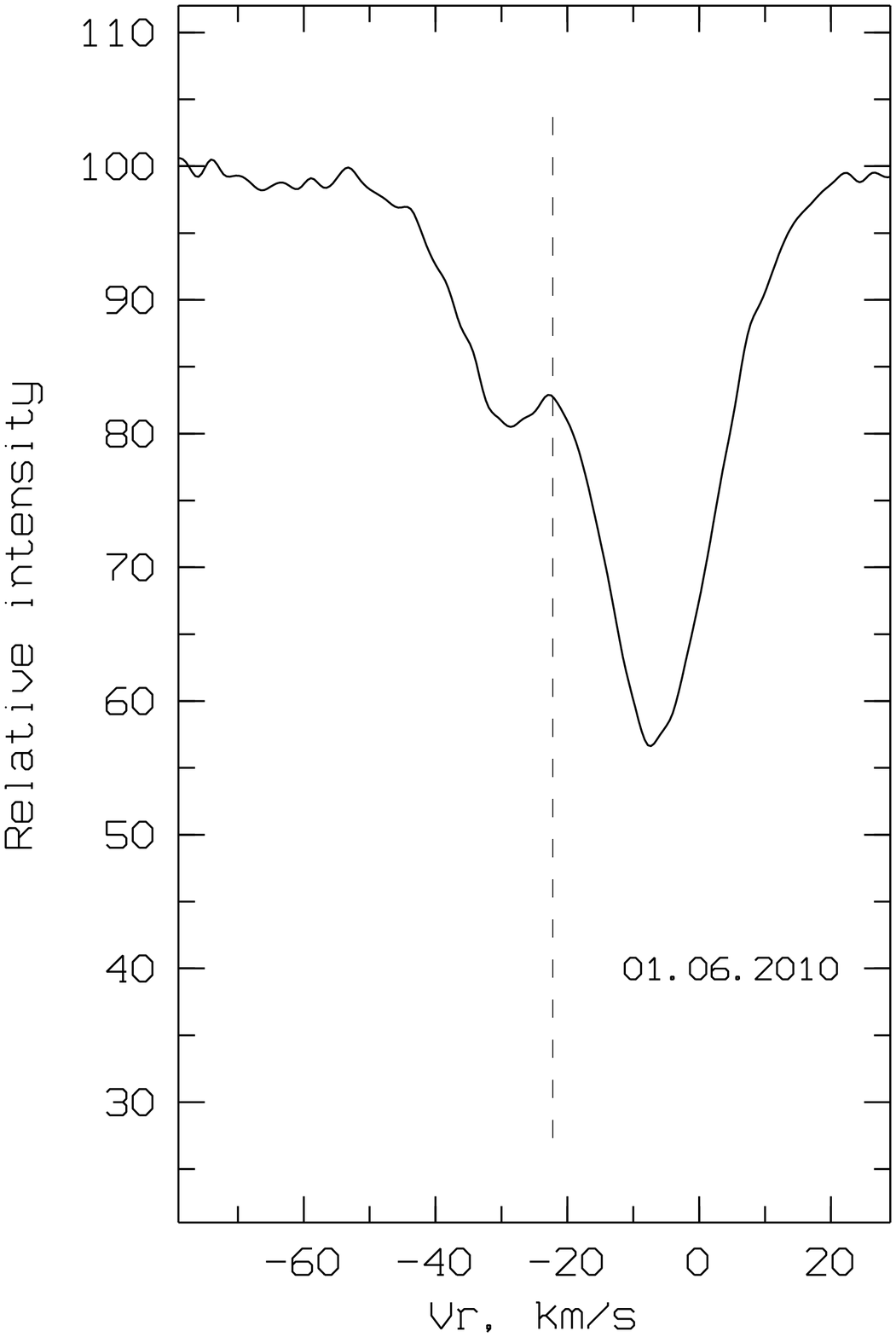}
\includegraphics[angle=0,width=0.35\textwidth,bb=40 60 540 790,clip]{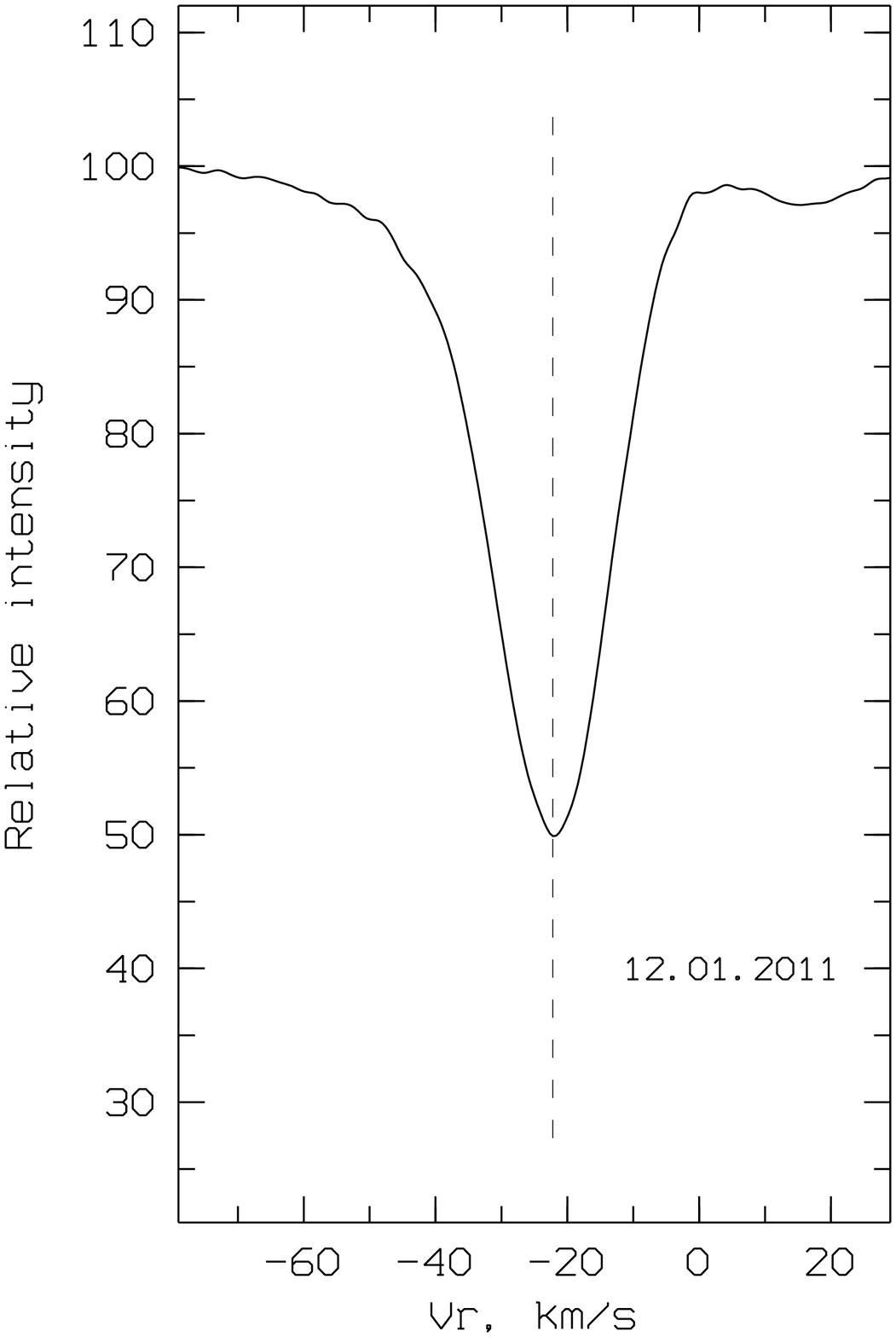}
\includegraphics[angle=0,width=0.35\textwidth,bb=40 60 540 790,clip]{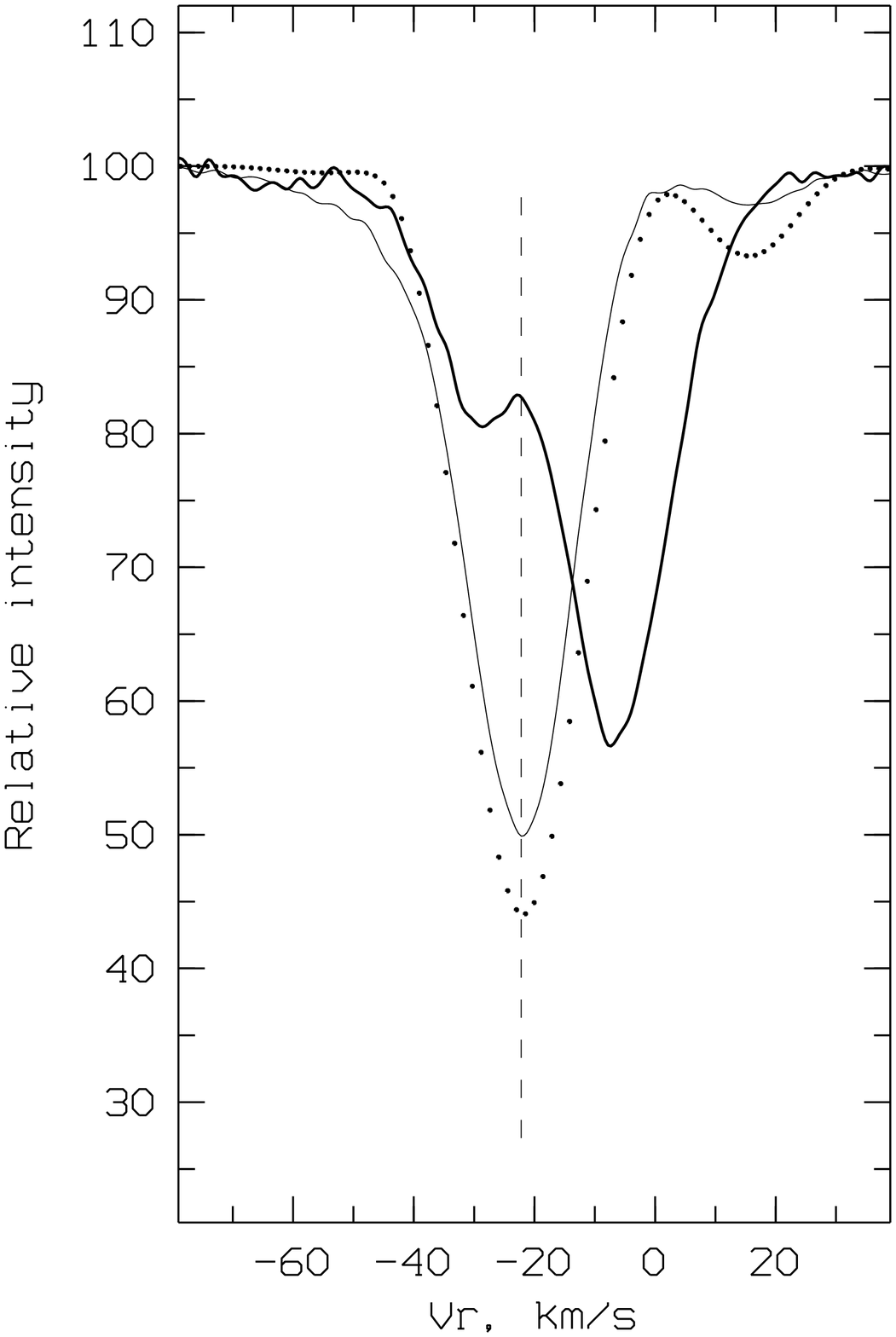}}
\caption{Profiles of the BaII\,6141\,\AA{} line in the spectra of LN\,Hya
       from various dates. The bottom right panel compares the observed
       profiles for 1 June, 2010 (thick curve) and January 12, 2011 (thin
       curve) to the theoretical profile (points) for Teff\,=\,6000\,K,
       log\,g\,=\,1.0, and $\xi_t = 4.7$\,km/s. The vertical dashed lines
       mark the adopted systemic velocity, Vsys\,=$-$21.6\,km/s.}
\label{Ba6141}
\end{figure}

\begin{figure}[t]	      		      
\hbox{
\includegraphics[angle=0,width=0.35\textwidth,bb=40 60 540 790,clip]{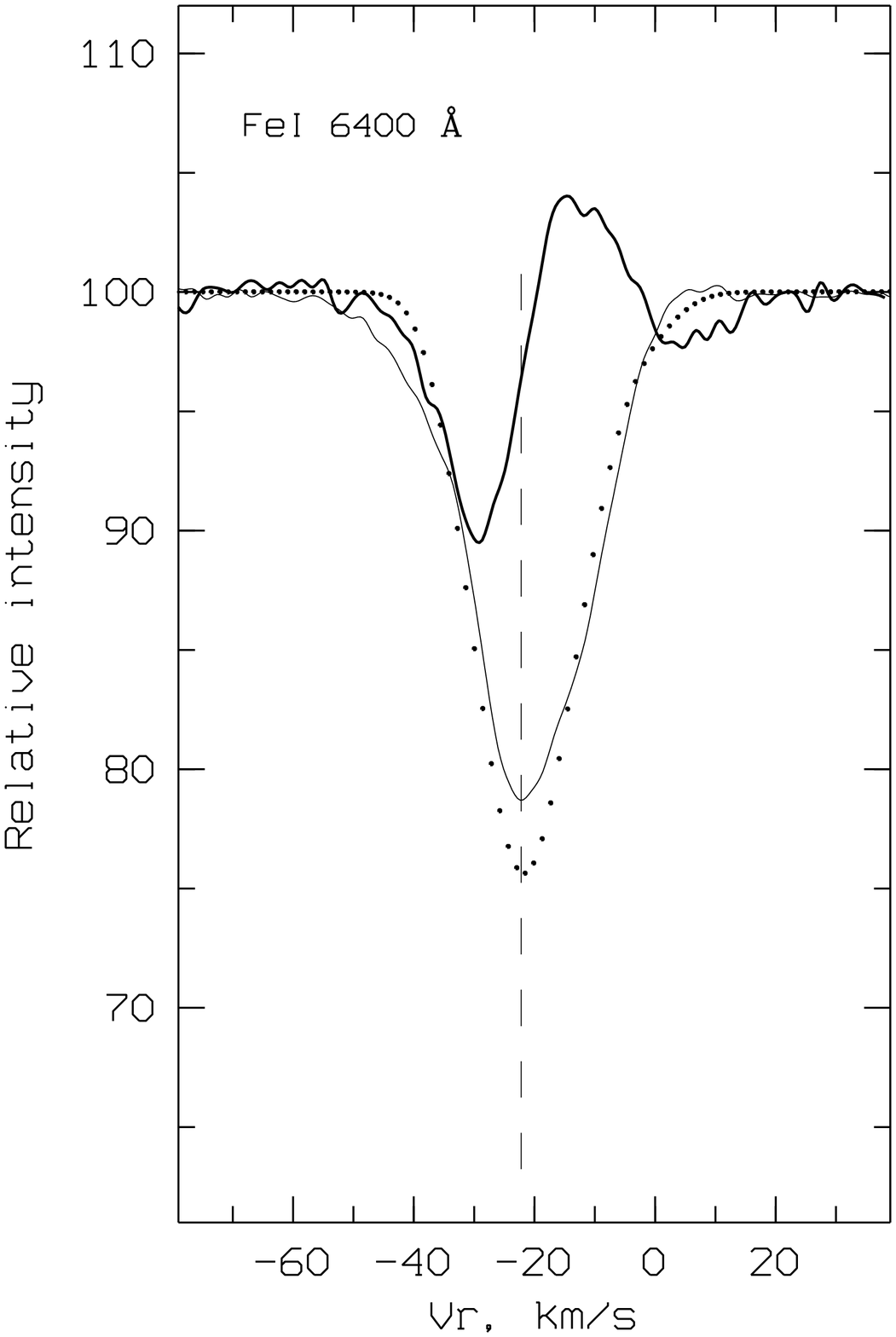}
\includegraphics[angle=0,width=0.35\textwidth,bb=40 60 540 790,clip]{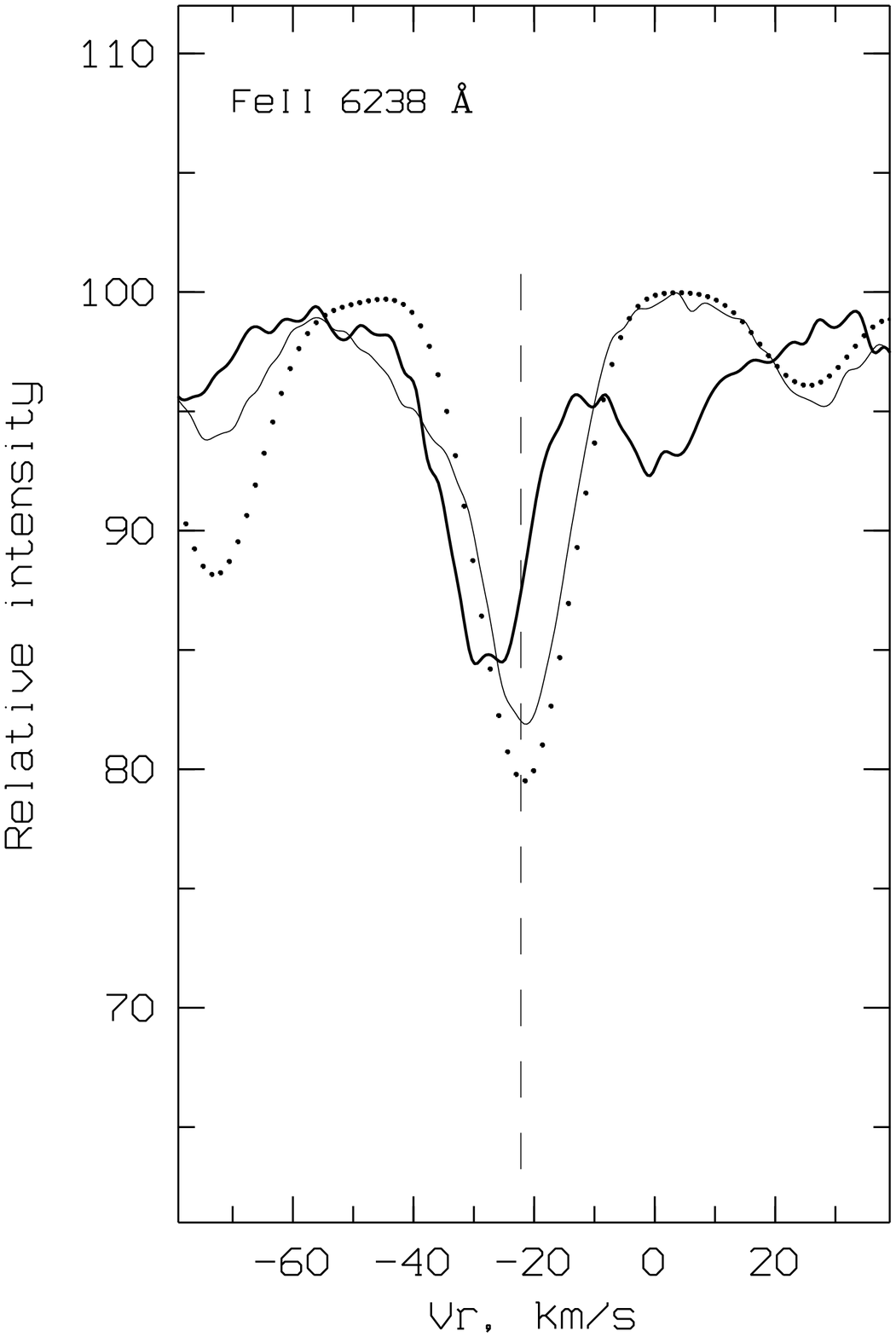}
\includegraphics[angle=0,width=0.35\textwidth,bb=40 60 540 790,clip]{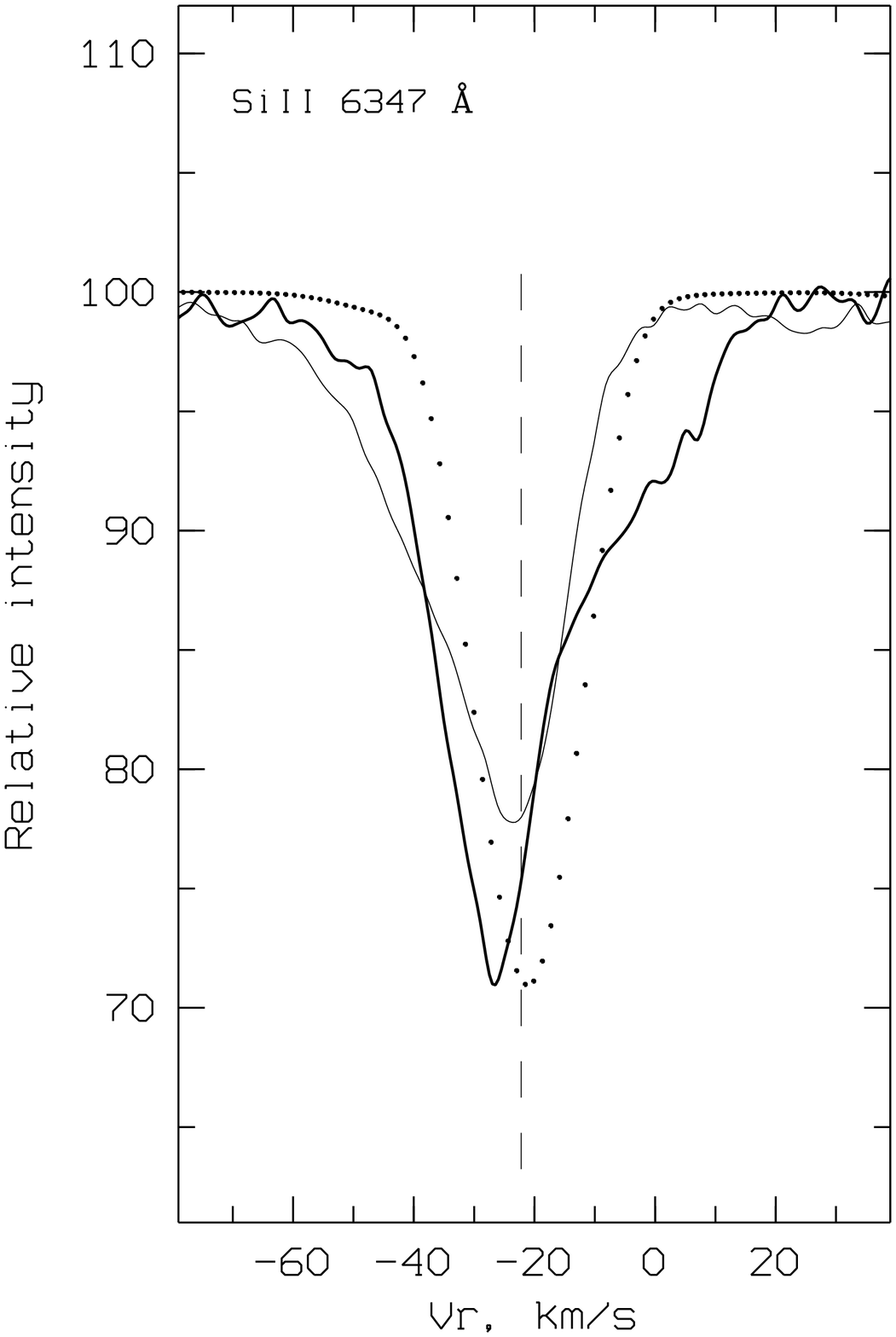}}
\hbox{
\includegraphics[angle=0,width=0.35\textwidth,bb=40 60 540 790,clip]{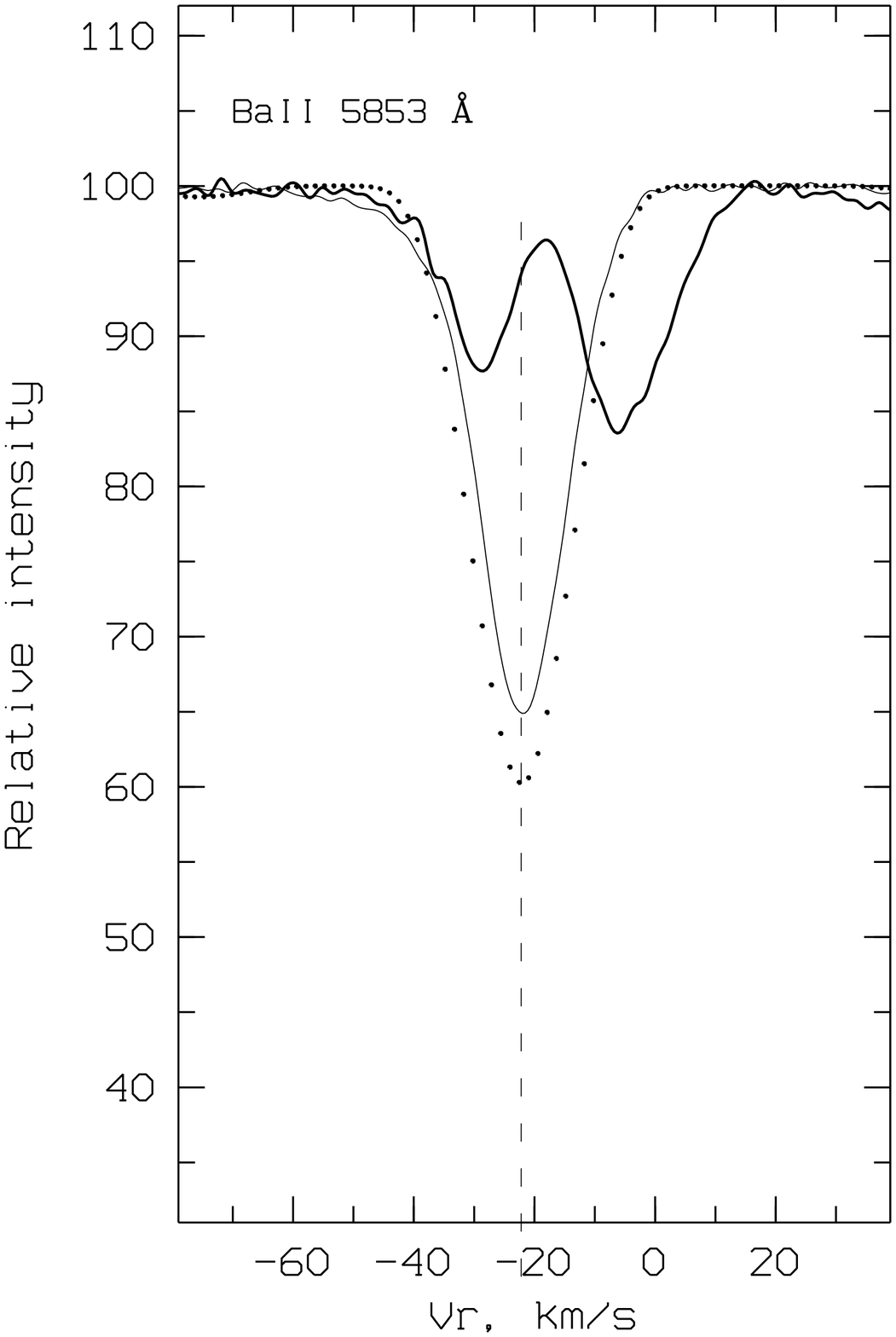}
\includegraphics[angle=0,width=0.35\textwidth,bb=40 60 540 790,clip]{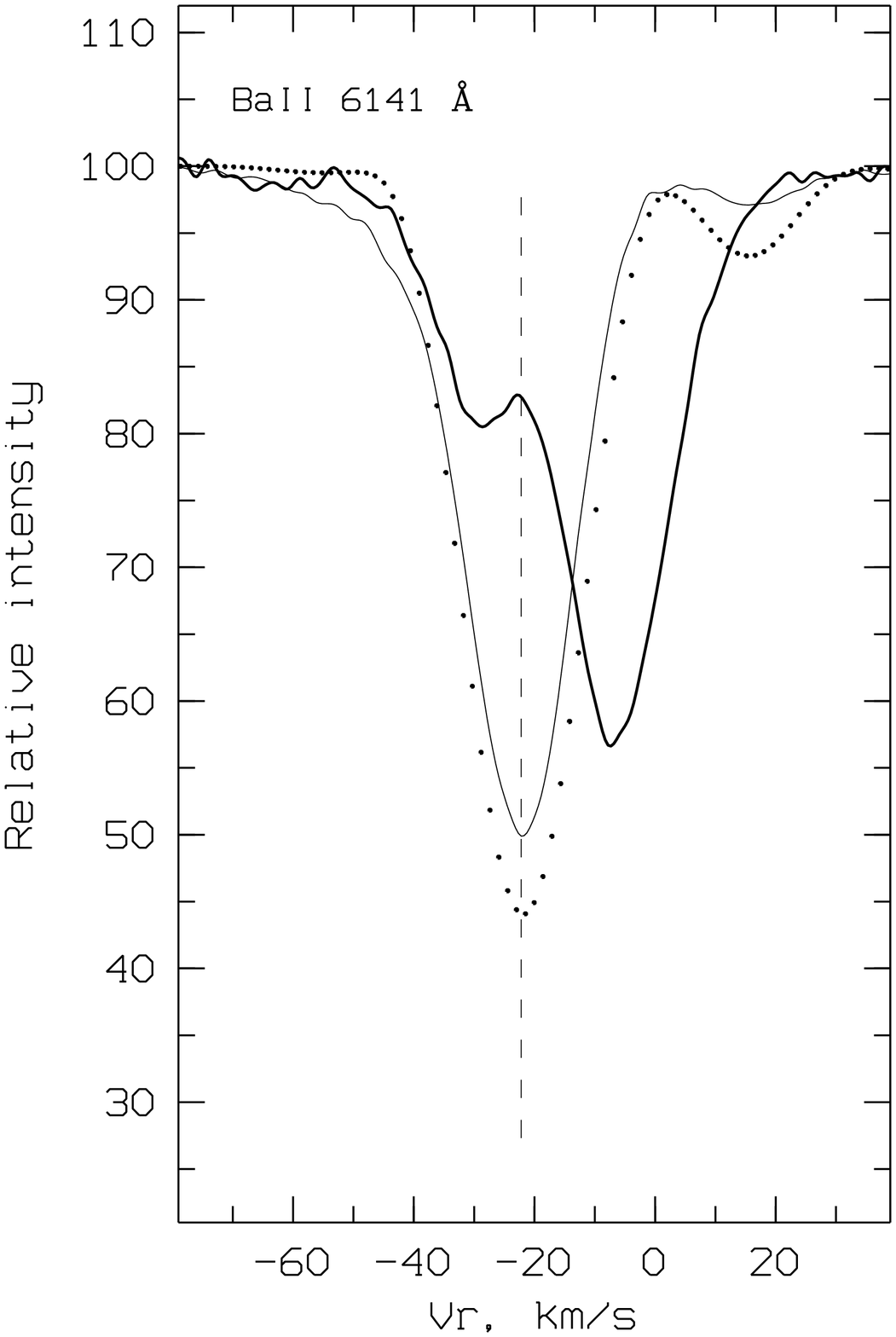}
\includegraphics[angle=0,width=0.35\textwidth,bb=40 60 540 790,clip]{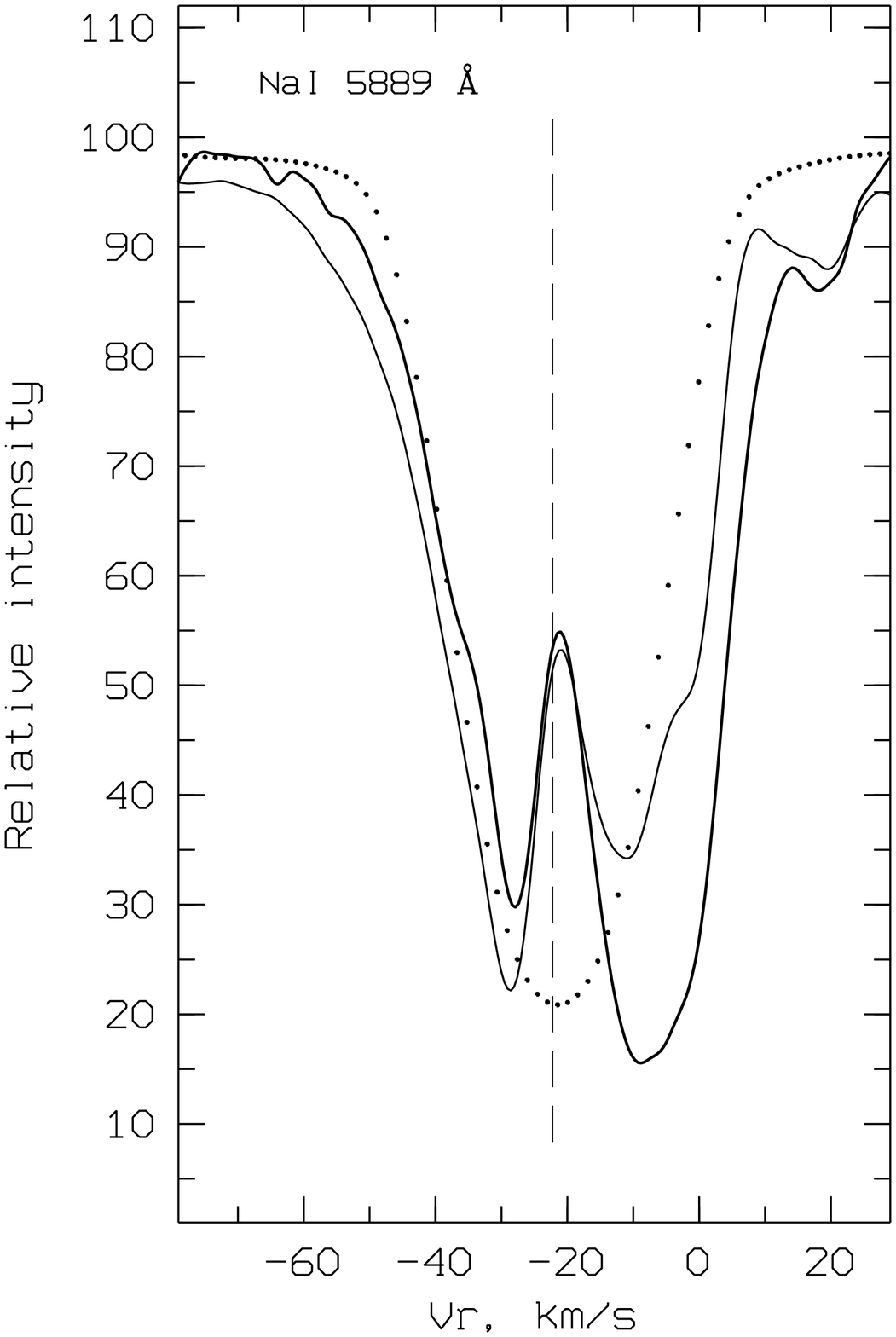}}
\caption{Profiles of selected lines in the spectra of LN\,Hya, compared
        to the theoretical profiles (points) for Teff\,=\,6000\,K, log\,g\,=\,1.0,
        and $\xi_t = 4.7$\,km/s. The thick and thin curves show the observed
        profiles for 1~June, 2010 and 12~January, 2011, respectively.}
\label{profiles}
\end{figure}

{\bf Strong absorption lines.} Apart from the changed H$\alpha$ profile,
analysis of our spectra of LN\,Hya taken in 2010$\div$2011 also reveals
previously unknown properties of the star's spectrum. A comparison of the
profiles and parameters of the spectral lines of LN\,Hya for different
nights demonstrates that the strongest metallic absorption lines, formed
in high layers of the stellar atmosphere, have complex timevariable
profiles. In the spectra obtained during 2~April~$\div$1~June, 2010, all
strong absorption lines in the yellow (FeII, TiII, SiII, and especially
BaII) have abnormal profiles. Figure\,\ref{Ba6141} presents profiles of the
BaII\,6141\,\AA{} lines for different observing epochs. The core of this
line in the spectrum for 21~February, 2003 was almost symmetric, but
changes were observed beginning from 2~April, 2010, which are most clearly
expressed in the spectrum for 1~June, 2010. The short-wavelength wings of
the strong absorption lines are lifted to some extent by emission lines,
and the absorption cores are sharpened or split due to the presence of
emission. As an example, Fig.\,\ref{profiles} presents profiles of the
Si\,II, Fe\,I, Fe\,II, Ba\,II, and D2~NaI\,(1) lines in the quiescent
(12~January, 2011) and most active (1~June, 2010) states. For comparison,
this figure also shows theoretical profiles for the parameters
Teff\,=\,6000\,K, log\,g\,=\,1.0, and $\xi_t = 4.7$\,km/s, which we
adopted in accordance with [16, 17].

There is obviouisly a contribution from upper atmospheric levels, which
provides the specific profile shapes. Several lines with distorted
profiles are listed in Table\,\ref{Asym}. In the last three columns of the
table, ``+'' denotes the presence of distortion for a given line, ``--''
an absence of distortion and ``+:'' line profiles without obvious emission
but with extended short--wave wings.

\begin{table}[hbtp]
\caption{Lines with asymmetric profiles in the spectra of LN\,Hya for
         various dates. In the last three columns of the table,
         ``+'' denotes the presence of distortion for a given line,
	 ``--'' an absence of distortion and ``+:'' line profiles without
	 obvious emission but with extended short--wave wings.}
\begin{tabular}{r l l|  c|  c|  c}
\hline
&  & &\multicolumn{3}{c}{Presence of asymmetry}\\  
\cline{4-6}
\multicolumn{2}{c}{Line} &$\chi_{low}$, eV &   05.04.2010 & 01.06.2010  &12.01.2011   \\  
\hline     
5167.33& MgI &2.71&+:  &   &  \\ 
5169.03& FeII&2.89&+   &   &  \\ 
5172.70& MgI &2.71&+   &   &  \\ 
5183.62& MgI &2.71&+   &   &  \\ 
5200.41& YII &0.99&+   &   &  \\ 
5232.94& FeI &2.94&+   &+  &+:  \\ 
5234.62& FeII&3.22&$-$ &+  &+:  \\ 
5237.33& CrII&4.07&$-$ &+  &+:  \\ 
5256.94& FeII&2.89&+   &+  &$-$ \\ 
5316.62& FeII&3.15&+   &+  &+:  \\ 
5336.77& TiII&1.58&+   &+  &+: \\ 
5371.49& FeI &0.96&+   &+  &+:  \\ 
5381.01& TiII&1.57&+   &+  &+:  \\ 
5393.17& FeI &3.24&+:  &+  &$-$  \\
5397.13& FeI &0.92&+   &+  &$-$ \\ 
5405.77& FeI &0.99&+   &+  &$-$  \\
5414.07& FeII&3.22&$-$ &+  &$-$ \\ 
5429.70& FeI &0.96&+   &+  &+:  \\ 
5446.92& FeI &0.99&+   &+  &+:  \\ 
5455.61& FeI &1.01&+   &+  &+:  \\ 
5526.79& ScII&1.77&+   &+  &+:  \\ 
5853.67& BaII&0.60&+   &+  &$-$  \\
6141.71& BaII&0.70&+   &+  &+:  \\ 
6245.62& ScII&1.51&+   &+  &$-$  \\
6247.55& FeII&3.89&$-$ &+  &$-$  \\
6347.09& SiII&8.12&+:  &+  &+:  \\ 
6371.36& SiII&8.12&+   &+  &+:  \\ 
6393.60& FeI &2.43&$-$ &+  &+:  \\ 
6400.00& FeI &3.60&+:  &+  &+: \\ 
6456.38& FeII&3.90&+:  &+  &+: \\ 
6496.90& BaII&0.60&+   &+  &+:  \\ 
\hline 
\end{tabular}
\label{Asym}
\end{table}

Abnormal P\,Cygni profiles in the spectrum of LN\,Hya are also exhibited by
lines of neutral iron. This efect is clearly expressed, for instance, in
the Fe\,I $\lambda$\,6400\,\AA{} line (top left panel of
Fig.\,\ref{profiles}); the excitation potential of the lower level of this
line ($\chi_{low} > 3$\,eV) is considerably higher than for the Ba\,II
lines ($\chi_{low} <1$\,eV). The Fe\,I\,$\lambda$\,6400\,\AA{} line is not
the only one in the spectrum that displays this property. Similar profiles
are also found for several other low--excitation lines of neutral iron and
calcium: Fe\,I\,$\lambda$\,6065.48\,\AA{},
Ca\,I\,$\lambda$\,6162.18\,\AA{}, Fe\,I\,$\lambda$\,6191.56\,\AA{},
Ca\,I\,$\lambda$\,6449.81\,\AA{}. The lower levels of these lines have
excitation potentials $\chi_{low} < 3$\,eV. Profiles of high-excitation
lines, such as Si\,II\,$\lambda$\,6347 or 6371\,\AA{} ($\chi_{low} >
8$\,eV), are less distorted. As an example, Fig.\,\ref{profiles} shows the
Si\,II\,$\lambda$\,6347\,\AA{} line profiles for two observing epochs.

\begin{figure}[t]	      		      
\includegraphics[angle=0,height=0.5\textwidth,width=0.6\textwidth,bb=40 60 540 790,clip]{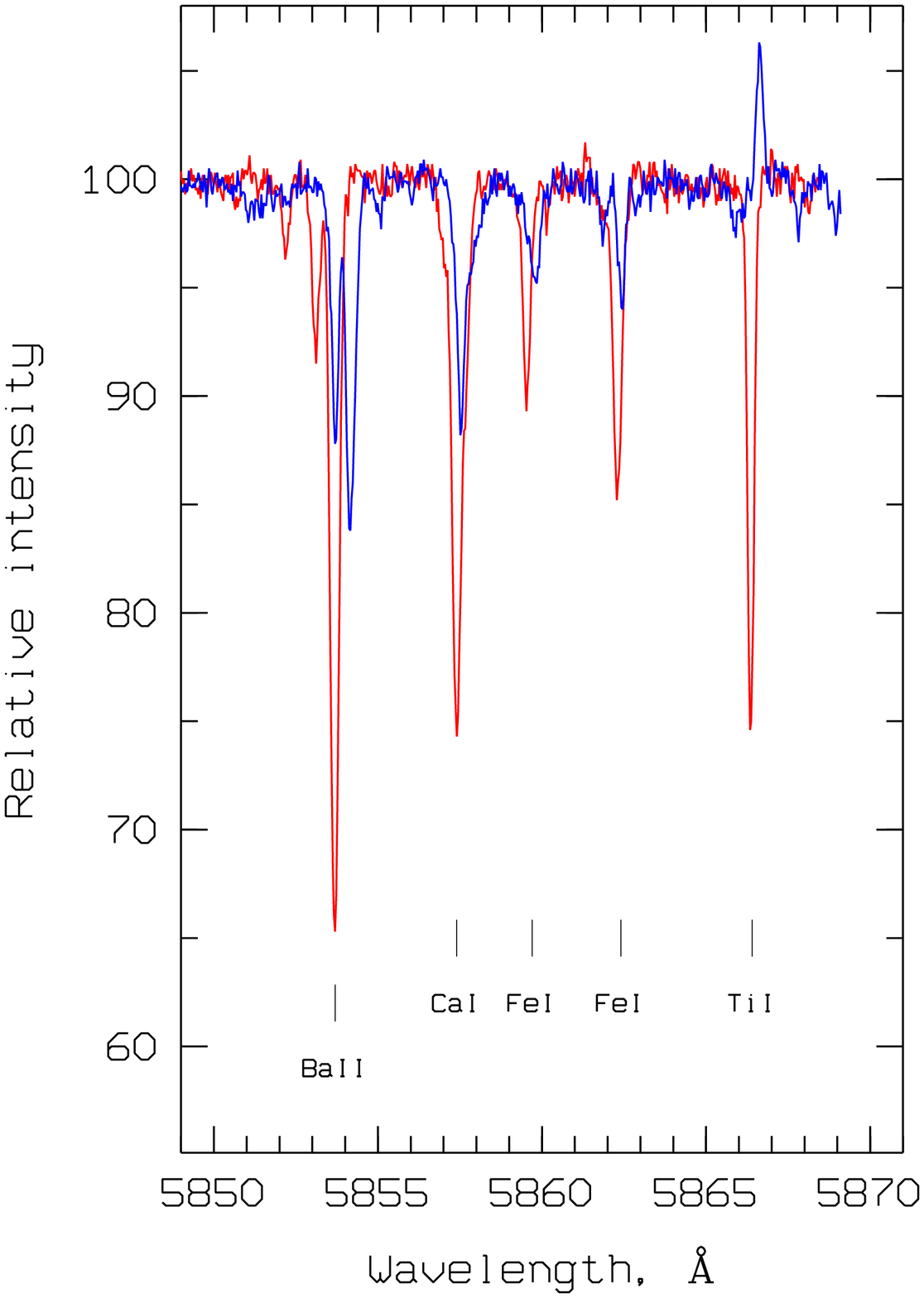}
\caption{Portions of the spectra of LN\,Hya with Ti\,I\,$\lambda$\,5866.40\,\AA{}
         emission, taken during quiescent (2~April, 2010; red curve) and excited
         (1~June, 2010; blue curve) atmospheric states. Identifications are
         presented for the main lines in this part of the spectrum.}
\label{Ti_emis}
\end{figure}

{\bf Emission Lines of Neutral Atoms.} The spectrum taken on 1~June, 2010
displays another previously unknown property of LN\,Hya: weak emission
lines of neutral atoms (V\,I, Mn\,I, Co\,I, Ni\,I, Fe\,I) with intensities
of several percent of the continuum. Table\,\ref{list_emis} presents a
list of these features, along with the excitation potentials of their
lower levels $chi_{\rm low}$ and radial velocities corresponding to the
positions of emission lines in three spectra. These emission features were
not present in spectra taken prior to June 2010. Instead, earlier spectra
displayed fairly strong absorption lines for the same atomic transitions.
As an example, Fig.\,\ref{Ti_emis} shows a fragment with one such emission
line, Ti\,I$\lambda$\,5866.40\,\AA{}. The spectrum of LN\,Hya taken on
1~June, 2010, during an ``excited'' state of the atmosphere, contains
Ti\,I\,$\lambda$\,5866.40\,\AA{} emission with an intensity of about 6\%,
while the quiescent atmosphere (2~April, 2010) exhibits absorption whose
position corresponds to the radial velocity derived from other undistorted
absorption lines in the same spectrum. The intensities of the emission
lines decrease in the 2011 spectra, and some disappear, as is demonstrated
by Table\,\ref{list_emis}.

The profiles of the D--lines Na\,I resonance doublet also display a
specific split shape in all our observations. A contribution from
interstellar Na\,I features is ruled out because of the star's high
Galactic latitude. In our opinion, the observed splitting of the Na\,I
lines is most likely due to the presence of emission formed in the
circumstellar medium, which splits each photospheric D--line into two
components (Fig.\,\ref{profiles}).

Thus, in the course of about 60~days between April and June 2010, we
detected the appearance and rapid increase of instability in the
atmosphere of LN\,Hya. The star's atmosphere had returned to its normal
state by the time of our most recent spectra (12~January, 2011 and
14~March, 2011), when the stellar spectrum, including the H$\alpha$
profile, was virtually undistinguishable from our first spectrum of
21~February, 2003.

\subsection{Radial velocity}

As was already noted, there were no earlier detailed radial-velocity
studies for LN\,Hya. The SIMBAD database presents only the mean (from three
heterogeneous measurements) heliocentric velocity of the star,
Vr\,=$-$22\,km/s. Thus, one of our tasks was to measure the pattern shown
by the radial velocities in detail, based on various spectral features, and
to study the time behavior of the velocity field. We measured the
heliocentric radial velocities Vr by moving the direct and mirror images
of the line profile until they coincided. This method can be used to
obtain Vr measurements for individual features in the profile. We compiled
a list of lines for Vr measurements using the spectral atlas [31],
prepared using similar (6-m telescope + NES) spectra of the post--AGB star
CY\,CMi (HD\,56126), whose atmospheric parameters and metallicity [10] are
close to those of LN\,Hya. The results of our Vr measurements for the six
available spectra are collected in Table\,\ref{date}, whose columns
present: (1) the observation date, (2) the corresponding Julian date JD,
(3) the recorded spectral range, (4) the heliocentric radial velocity
Vr(metals) averaged over a large set of symmetric metallic absorptions,
with the number of lines at the measured positions indicated in brackets,
and (5) the velocities obtained from the H$\alpha$ absorption component
and the emission component of the D-lines of the Na\,I doublet.

Unfortunately, the systemic velocity of LN Hya was not determined.
Attempts were made to search for spectral features that could have been
formed in the stellar shell (emission bands of OH, H$_2$O masers, etc.),
but without success. In particular, Venn et al. [32] observed a sample of
AGB and post--AGB stars in the submillimeter, but did not detect any flux
in the $^{12}$CO lines for LN\,Hya, which is not a surprise for an O--rich
star. We adopted the velocity we measured for weak symmetric emission
lines of neutral atoms (V\,I, Mn\,I, Co\,I, Ni\,I, Fe\,I) as the systemic
velocity. Table\,\ref{list_emis} shows that the mean radial velocity for
these features, Vsys\,=$-21.6$\,km/s, is stationary in time. We derived
the same velocity (Vr\,=$-21.6$\,km/s) from the emission component of the
Na\,I D--lines (see the last column of Table\,\ref{date}), and this value
likewise does not change with time. These two findings suggest that the
Na\,I~D emission is formed in the circumstellar structure.

The line-profile anomalies in the spectrum of LN Hya influence the
radial velocities measured from these lines. The derived velocities depend
on the type of line and the particular features used for the measurements,
as well as the applied technique. Most metallic lines in the spectra of
LN\,Hya are asymmetric: the blue wing is either extended (see the spectrum for
21~February, 2003, where the velocity measured from the upper part of the
profile is lower than the velocity measured from the core) or distorted
by emission (the spectra of April and June 2010, where the velocity
measured from the upper part of the profile is higher than the velocity
measured from the core). To reveal possible time variations of Vr, we
selected only symmetric absorption lines of metals without visible
distortions in each spectrum.

Our analysis of the Vr sets obtained from symmetric absorption lines led
to the conclusion that there was no relation between the intensity and the
corresponding radial velocity. This means we can use the average radial
velocity for each spectrum in our further analysis. The calculated mean
velocity Vr(metals) is presented in the fourth column of
Table\,\ref{date}. The velocity Vr(metals), reliably determined from
numerous symmetric absorption lines, changes from epoch to epochr with a
small amplitude of 2$\div$3\,km/s. These low-amplitude Vr variations are
clearly due to pulsations of the stellar atmosphere. This conclusion
agrees with the brightness variability of LN\,Hya ($\Delta$B\,=\,0.32$^m$)
[1]. Vr variations with a similar amplitude were recently detected by
Takeda et al. [18], who measured heliocentric radial velocities Vr from
$-25.9$ to $-31.1$\,km/s in four high-resolution spectra (R\,=\,70000)
taken within a week.

The profile variations of strong absorption lines in the spectra of
LN\,Hya are due to variations in the shifts of absorption lines and the
presence of an emission component whose position does not change. The
combination of these two factors splits the absorption into components
whose positions and intensities vary. The radial velocity derived from the
H$\alpha$ absorption component varies only slightly. Apart from the epoch of
strongest atmospheric excitation (1~June, 2010), the mean velocity for
four phases is Vr\,=$-22.3$\,km/s, close to the mean velocity indicted by
symmetric absorption lines of metals.

\begin{table}[h]
\bigskip
\caption{Radial velocities Vr(emis) measured for LN Hya on three dates from
        symmetric emission lines of neutral atoms (the last line contains the
        mean velocity for the corresponding date)}
\begin{tabular}{l r|  c|  c| r}
\hline
&  & \multicolumn{3}{c}{Vr(emis), km/s}\\  
\cline{3-5}
Line  &$\chi_{low}$, eV  &01.06.2010  &12.01.2011  & 14.03.2011 \\  
\hline     
5394.68 MnI& 0.00& $-20.32$& $-21.97$   &            \\ 
5490.15 TiI& 1.46& $-21.16$& $-23.02$   &            \\ 
5644.14 TiI& 2.27& $-23.30$& $-21.44$   &$-21.69$    \\ 
5727.05 VI & 1.08& $-21.39$& $-19.51$   &$-21.05$    \\ 
5866.40 TiI& 1.07& $-22.14$& $-22.06$   &$-22.49$    \\ 
5918.55 TiI& 1.06&         & $-19.86$   &            \\ 
5953.16 TiI& 1.89& $-21.91$&            &            \\ 
5956.70 FeI& 0.86&         &            &$-21.05$    \\ 
6091.18 TiI& 2.27& $-22.98$&            &            \\ 
6108.12 NiI& 1.67& $-21.15$& $-23.02$   &            \\ 
6126.22 TiI& 1.07& $-21.77$&$-21.77$    &$-19.59$    \\ 
6261.10 TiI& 1.43& $-20.62$&            &$-22.02$    \\ 
6450.24 CoI& 1.71& $-21.99$& $-21.54$   &$-23.15$    \\ 
\hline 
\multicolumn{2}{c|}{\small Vr(aver), km/s}&$-21.7$ & $-21.6$ &$-21.6$  \\   
\hline
\end{tabular}
\label{List_emis}
\end{table}

\section{Discussion}

\subsection{Pulsations and atmospheric kinematics of LN\,Hya}

The peculiarities of the H$\alpha$ profiles and the strong absorption lines
in the spectrum of LN\,Hya (the presence of emission that splits absorption
lines into components) are consistent with the existence of pulsations and
the passage of a shock in the atmosphere. In his well-known paper,
Schwarzschild [33] concluded that split absorption profiles and H$\alpha$
emission were signs of a shock passing through the stellar atmosphere.
This explanation seems fairly natural for a star with Teff\,=\,6000\,K and
log\,g\,=\,1.0. Unfortunately, we could not find any photometric data that
could contribute to a better understanding of the observed spectral
peculiarities for our observing epochs.

We observe signs of motions in deep atmospheric layers in both the
quiescent and active states, such as the differing radial velocities
Vr(metals) measured from numerous symmetric absorption lines of metals in
Table\,\ref{Asym}, and the presence of extended short-wave wings of strong
absorptions, which provide evidence for high-velocity outward motions.
Variable emission in both H$\alpha$ and strongest lines of metals is
observed in the active state (the spectra for 2010). If we restrict our
consideration to the H$\alpha$ profile, the observed pattern suggests a
shock moving in the atmosphere of a pulsating star. Subtracting the mean
profile for the quiescent spectra from the H$\alpha$ profile observed in
the active state yields an inverse P\,Cygni profile (see also the bottom
right panel of Fig.\,\ref{Halpha}). If we exclude the contribution from
the stable circumstellar structure, we observe matter falling onto the
star in the active state (in the H$\alpha$). However, considering other
spectral features, we find that the emission components of the metal lines
are formed in various outward--moving layers (such as
Ba\,II\,$\lambda$\,6141\,\AA{}), as well as back, towards the photosphere
(Fe\,I\,$\lambda$\,6400\,\AA{}, Fe\,II\,$\lambda$\,6238\,\AA{},
Ba\,II\,$\lambda$\,5853\,\AA{}). Thus, it seems difficult to relate the
formation of these emission lines to a single de-excitation region behind
the front of a single shock.

Aikawa [34, 35] demonstrated that, in the case of low-mass UU\,Her
supergiants, low-amplitude pulsations could appear when the luminosity was
sufficiently high, and the pulsations of higher-luminosity objects
could become irregular and even chaotic. Pulsations with an amplitude of
about 15\,km/s were detected in the atmosphere of the closest relative of
LN\,Hya the high-latitude supergiant UU\,Her, whose pulsation periods, P\,=\,73d
in the fundamental mode and P\,=\,45d in the first overtone, are
known from photometric data [36]. Based on multi-epoch spectroscopy of
UU\,Her, Klochkova et al. [28] noted asymmetry of the strongest ion lines
(Ba\,II, Si\,II), as well as a strong shift of the H$\alpha$ absorption core
relative to other absorption lines in each of the spectra. In the present
study, we have detected similar spectral properties for LN\,Hya, but the
pulsation amplitude is much lower than for UU\,Her.

In the active state, we observe another interesting property of LN\,Hya:
the absorption components of strong lines are shifted towards longer
wavelengths (Table\,\ref{date}). These shifts cannot be due to the
``cutting'' of the absorption--profile core by a short-wave emission
component, since the absorption wings are also shifted (see, for
example, the Ba\,II\,$\lambda$\,6141\,\AA{} profile in the 1~June, 2010
spectrum in Fig.\,\ref{Ba6141}). This supports the hypothesis that matter
is falling onto the star.

Thus, the active state of the atmosphere of LN\,Hya observed in 2010
was primarily determined by the motion of matter towards the star (the
inverse P\,Cygni profile of the H$\lambda$ lines and redshifted cores of
strong lines). This motion coexists with outward motions (the extended
short-wave wings of metal lines). Simultaneously, we observe two
regions of de-excitation in the metal lines, with opposite motions
relative to the stellar center. The active atmospheric state is
accompanied by the excitation of neutral-atom emission lines at the
periphery of the circumstellar shell. In order for these phenomena to be
consistent, we must reject the hypothesis of spherical symmetry of the
structure of the atmosphere and circumstellar shell.

Photometric manifestations of semiregular variability depend appreciably
on the fraction of the absorption spectrum formed in the shell. For
example, the extended atmospheres of RV~Tauri stars contain enough matter
for division into two or three absorption systems to become visible with
sufficiently good spectral resolution. Ways to reconcile the two-peaked
light curve of AC~Her with the single-peaked curve of the equivalent-width
variations for strong lines of metals, mainly formed in the shell, are
discussed in [37].

The abrupt change in the parameters of spectral features we have detected
suggests that the critical state of the atmosphere of LN\,Hya during the
first half of 2010 could be due to a change in the pulsation period.
This behavior is characteristic of supergiants at high Galactic latitudes
[38]. Confirmation of this hypothesis would obviously require long-term
spectroscopic monitoring of the star with good time coverage.

\subsection{Pulsations of related objects}

Pulsation instability is characteristic of many post-AGB objects, in
agreement with theoretical computations of the pulsations of stars
evolving from the AGB to planetary-nebula phase [34, 35, 39]. The physical
mechanism exciting pulsations of post--AGB stars, which consist of a
degenerate CO core (with a typical core mass of 0.6${\mathcal M}_{\odot}$)
surrounded by an extended, low-mass atmosphere, remains unknown. Unstable,
low-mass ($0.6\div 0.8{\mathcal M}_{\odot}$), high-luminosity ($0.0 \le
\log g \le 1.8$) post--AGB stars that pulsate with low amplitudes in high
modes populate an instability strip in the Hertzsprung--Russell diagram
at temperatures of 5010\,K$\le$\,Teff\,$\le$7940\,K, shifted relative to
the classical instability strip towards hotter stars [40]. Such
low-amplitude pulsations have been successfully modeled. Fokin et al. [41]
calculated grids of models reproducing irregular brightness and
Vr variations and the corresponding power spectrum for
post--AGB stars over wide intervals of mass, effective temperature, and
luminosity.

Multi-component structures of metal lines due to passing shocks are
observed in the spectra of various types of variable stars, including
classical Cepheids (e.g., [42]) and RV~Tauri pulsating stars [43]. No H$\alpha$
emission is detected in the spectra of classical Cepheids [44]. A
combination of two phenomena --  split absorption lines and H$\alpha$
emission -- is observed in the spectra of Population~II pulsating stars
(see [44] and references therein). A good example is the halo pulsating
star W\,Vir. Recently, Kovtyukh et al. [45] used spectroscopic monitoring
of this Pop\,II Cepheid to study the time behavior of the H\,I and He\,I
emission profiles, split metal lines (Fe\,I, Fe\,II, Ba\,II, Na\,I, etc.), and
the velocity field. They suggested a schematic model of the system
(star\,+\,shell) that can explain time variations of the velocity field
in the stellar atmosphere with the passage of two shocks [45]. It is more
difficult to study pulsations in post--AGB objects with strong
circumstellar shells. Due to the action of complex dynamical processes and
the lack of the required quantity and quality of observations, atmospheric
pulsations of stars evolving beyond the AGB stage remain poorly studied.

Sufficiently high-quality spectroscopic monitoring (high spectral
resolution and low noise) over a wide wavelength range has been performed
for only one of the brightest post--AGB stars, CY\,CMi. This is a rare star
at this evolutionary stage, with products of the third dredge--up detected
in its atmosphere [10]. Lebre et al. [46] used Fourier analysis to study
H$\alpha$ variations and the corresponding set of radial velocities,
concluding that the dynamics of the CY\,CMi atmosphere are complex due to
pulsations. Barthes et al. [47] obtained a large series of good--quality
spectroscopic observations of CY\,CMi over almost eight years. To study the
Vr variations and search for a pulsation period, they measured the
positions of two symmetric lines, Ba\,II\,$\lambda$\,5853 and
C\,I\,$\lambda$\,6587\,\AA{}. They found a first-overtone pulsation
period, P\,=\,36.8d, with an amplitude of $\Delta$Vr\,=\,2.7\,km/s based
on 89 spectroscopic epochs. A conclusion of [47] was that the observed
variations of the neutral-hydrogen lines (H$\alpha$ and H$\beta$) could
not be due to pulsations.

Studies of velocity fields for post-AGB stars are often complicated by
the presence of combined variations of Vr due to pulsations and
differential motions in the extended atmospheres of semiregular
variable stars. Klochkova and Chentsov [48] used (6m telescope + NES) spectra of
CY\,CMi to study the radial velocities in its atmosphere and shell. Their
high quality observations over a wide wavelength range enabled them to
study a large set of lines with various intensities, and to detect
considerable differential line shifts within a single spectrogram, as
well as changes of these shifts with time. They concluded that both
expanding and infalling layers were simultaneously present in the
atmosphere of CY\,CMi [48].

Equally complex kinematic patterns are observed for two variable stars
with similar properties, V354\,Lac and V448\,Lac. The cool variable
V354\,Lac, identified with the infrared source IRAS\,22272+5435, is one of
the most interesting post--AGB stars. It was one of the first post-AGB
stars with a 21\,$\mu$ spectral feature, and 6-m telescope spectra
revealed large excesses of carbon and $s$-process elements in the atmosphere
[49]. Later, optical spectra of V354\,Lac obtained with high spectral
resolution using the 6m telescope\,+\,NES instrument in 2001$\div$2008
showed that the strongest absorption lines with lower-level excitation
potentials below 1\,eV were split [7, 8]. Analysis of the kinematic pattern
demonstrated that the shortwavelength component of the split line was
formed in the star'ss strong gas and dust shell. Profile variations
of strong absorption lines were detected.

Multiple high-spectral-resolution observations in 1998$\div$2008 were used
to study time variations of spectral profiles and the velocity field in
the atmosphere and circumstellar shell of the semiregular variable
V448\,Lac [9]. Like those of LN\,Hya and V354\,Lac, spectra of V448\,Lac
reveal asymmetry and profile variability of the strongest absorption lines
with lower-level excitation potentials $\chi_{low} < 1$\,eV, first and
foremost, for the profiles of the strongest lines of the Ba\,II, Y\,II,
La\,II, Si\,II ions. The peculiarity of these profiles can be explained by
a superposition of stellar absorption lines and shell emission lines.
Emission of the Swan~(0;1)\,5635\,\AA{} band of the C$_2$ molecule was
detected for the first time in the spectrum of V448\,Lac.
Vr variations with an amplitude of about 8\,km/s were found
for the H$\alpha$ core. Lower-amplitude (about 1$\div$2\,km/s) Vr variations for
weak metallic absorption lines may reflect pulsations in the atmosphere.
Differential shifts of the lines from 0 to 8\,km/s were measured at
different epochs. The molecular spectrum is stationary in time,
indicating a constant expansion rate of the circumstellar shell, as
measured from C$_2$ and NaI lines: Vexp\,=\,15.2\,km/s.

Finally, further studies of post--AGB stars, including spectroscopic
monitoring and spectropolarimetry, are needed to improve our understanding
of the detected variations of Vr and the spectralline profiles of these
stars and develop models for such systems. High-spectral resolution
spectropolarimetry will provide key information about the structure of the
circumstellar environment. Long-term spectroscopic monitoring can reveal
secular variations of atmospheric parameters and distinguish genuine
post-AGB stars from objects mimicking this stage [50]. Another advantage
of many-year monitoring is the possibility of detecting smallamplitude
radial-velocity variations that may indicate spectroscopic binarity. 

\section{Conclusions}

Our echelle spectra of the low-mass, high-latitude supergiant
LN\,Hya acquired with the 6-m telescope over several years over a wide
spectral range and with high spectroscopic resolution have revealed new
spectroscopic phenomena for this star, and enabled us to measure radial
velocities for spectral features formed at various depths in the stellar
atmosphere.

Variations in the radial velocities measured from weak photospheric
absorption lines indicate the presence of low-amplitude pulsations in deep
(nearly photospheric) layers of the atmosphere.

We have found complex, time-variable profile shapes for strong absorption
lines (Fe\,I, Fe\,II, Si\,II, Ba\,II, etc.) formed in upper layers of the
star's extended atmosphere. We have detected variations of the asymmetries
of these absorption lines from spectrum to spectrum, including variations
in their splitting.

A previously unknown property of LN\,Hya appeared in the spectrum of
1~June, 2010: weak symmetric emissions identified with lines of
neutral atoms (V\,I, Mn\,I, Co\,I, Ni\,I, Fe\,I). The time--stationary mean
Vr from these emission lines, formed in the circumstellar
medium (Vsys\,=$-21.6$\,km/s), can be adopted as the systemic velocity.

During the 2010 observing season, the position and depth of the H$\alpha$
absorption component, intensities of the short-wave and
long-wave emission lines, and the intensity ratio of these lines
varied from spectrum to spectrum. These properties of the stellar
spectrum, detected for the first time in these observations, suggest
that we detected a rapid change of the physical conditions in the upper
atmospheric layers of LN\,Hya in 2010, possibly due to the passage of a
shock.

\section*{Acknowledgments}

The authors are grateful to M.V. Yushkin for his assistance during the
observations. This study was supported by the Russian Foundation for Basic
Research (projects 08--02--00072\,a and 11--02--00319\,a) and the Basic
Research Program of the Presidium of Russian Academy of Sciences ``The
Origin, Structure, and Evolution of Objects in the Universe''.

\end{document}